\documentclass[prd,showpacs,showkeys,nofootinbib,floatfix,eqsecnum,
               fleqn,preprint,12pt,tightenlines]{revtex4-1} 

\usepackage{amsfonts,amsmath,amssymb}
\usepackage{bm}
\usepackage{verbatim}
\usepackage{graphicx}
\usepackage{MnSymbol}  

\newcommand{\beq}{\begin{equation}}
\newcommand{\eeq}{\end{equation}}
\newcommand{\beqa}{\begin{eqnarray}}
\newcommand{\eeqa}{\end{eqnarray}}
\newcommand{\bsubeqs}{\begin{subequations}}
\newcommand{\esubeqs}{\end{subequations}}

\newcommand{\half}{{\textstyle \frac{1}{2}}}    
\def\id{\makebox[0.6ex][l]{$1$}{\rm l}}   

\newcommand{\Chi}{X}   

\begin{document}

\begin{widetext}
%
%
\noindent Class. Quant. Grav. \textbf{40} (2023) 124001
\hfill    arXiv:2212.00709
%
%
\newline\vspace*{3mm}
\end{widetext}

\title{Emergent gravity from the IIB matrix model and\\
cancellation of a cosmological constant}

\author{\vspace*{5mm} F.R. Klinkhamer}
\email{frans.klinkhamer@kit.edu}
\affiliation{Institute for Theoretical Physics,
Karlsruhe Institute of Technology (KIT),\\ 76128 Karlsruhe,  Germany\\}

\begin{abstract}
\vspace*{2mm}\noindent
We review a cosmological model where the metric determinant
plays a dynamical role and present new numerical results
on the cancellation of the vacuum energy density
including the contribution of  a cosmological constant.
The action of this model is only invariant under restricted
coordinate transformations with unit Jacobian
(the same restriction appears in the well-known unimodular-gravity
approach to the cosmological constant problem).
As to the possible origin of the nonstandard terms in the matter
action of the model, we show that these terms can, in principle,
arise from the emergent gravity in the IIB matrix model,
a nonperturbative formulation of superstring theory.
\end{abstract}

\maketitle

\section{Introduction}
\label{sec:Introduction}

The standard model of elementary particle physics
describes the electromagnetic and strong
interactions of the particles.
With $c=1$  and $\hbar=1$ from natural units,
the corresponding quantum field theory involves,
in the strong sector (quantum chromodynamics),
a vacuum energy density $\epsilon_{V}^\text{\,(QCD)}$
of the order of $(100\;\text{MeV})^4 \sim 10^{32}\;\text{eV}^4$
and, in the electroweak sector,
a vacuum energy density $\epsilon_{V}^\text{\,(EW)}$
of the order of $(100\;\text{GeV})^4 \sim 10^{44}\;\text{eV}^4$.
The astronomical observations, however, give a
cosmological constant $\Lambda$ which is of the order
of $10^{-11}\;\text{eV}^4$ (the corresponding vacuum energy density 
is $\rho_\text{vac}=\Lambda$ and the vacuum pressure
$P_\text{vac}= - \rho_\text{vac}=-\Lambda$).

The cosmological constant problem (CCP)
is about explaining how the huge vacuum energy densities
of elementary particle physics \emph{naturally} give rise to
the present Universe with a tiny value of the vacuum energy density,
where there are some 55 orders of magnitude to account for.
The various theoretical aspects of the
cosmological constant problem (CCP) are discussed
in, for example, Weinberg's review~\cite{Weinberg1989}.
The decisive astronomical observations of a
nonzero cosmological constant are reviewed
by Carroll~\cite{Carroll2001} and the
most recent observations are
covered in Chap.~28 of the Review of Particle Physics~\cite{RPP2022}.

As to the ``meaning'' of the cosmological constant $\Lambda$,
an interesting idea appears in the so-called
unimodular-gravity approach to the CCP, which goes
back to a 1919 paper by Einstein~\cite{Einstein1919}
and has resurfaced in more recent papers~\cite{vanderBij-etal1982,Zee1983,%
BuchmuellerDragon1988,HenneauxTeitelboim1989}.
Typically, the metric determinant is eliminated as a dynamical variable
and $\Lambda$ appears in the field equations
as an integration constant.
There is, however, no explanation of the actual
experimental value $\Lambda \approx (2\,\text{meV})^{4}$.

It appears that there have been many different contributions to the
vacuum energy density occurring over the whole history of the Universe
and some form of adjustment mechanism seems to be called for.
A particular type of adjustment mechanism has been proposed,
which is inspired by condensed matter physics.
In that approach, there is a special type of vacuum variable $q$,
which provides
for the natural cancellation of any previously generated
vacuum energy density~\cite{KlinkhamerVolovik2008a,KlinkhamerVolovik2008b}.
Several follow-up papers on the $q$-theory approach to the CCP
have appeared over the years~\cite{KlinkhamerVolovik2008c,%
KlinkhamerVolovik2009,KlinkhamerVolovik2009-gluonic,%
KlinkhamerVolovik2009-ew,KlinkhamerVolovik2016-4Dbrane,%
KlinkhamerVolovik-MPLA-2016,KlinkhamerVolovik2016-DM,%
KlinkhamerVolovik2019-tetrads,%
KlinkhamerVolovik2022-BBasTopQPT,Klinkhamer2022-preprint-4Dbrane}.
Here, we should perhaps emphasize one point,  
namely that $q$-theory in the cosmological context leads
to \emph{universal} dynamic equations, independent of
the particular realization of the $q$ variable
(see, in particular, App. A 3 of 
Ref.~\cite{KlinkhamerVolovik2022-BBasTopQPT}).

In nearly all previous work on $q$-theory, there was a postulated
field, from which the $q$ variable was obtained.
Recently, we have explored the idea of getting a $q$-type
field by use of the already available fields of
general relativity and the standard model, possibly
re-interpreting one or more of these fields.
It turns out that the metric determinant can play
the role of such a $q$-type field~\cite{Klinkhamer2022-ext-unimod}.
This makes the metric determinant a physical variable
and restricts the allowed coordinate transformations,
which brings us back to the unimodular-gravity approach
mentioned above. But with a difference:
in the unimodular-gravity approach,
the metric determinant can be removed altogether as
a dynamical variable, whereas, in our approach,
the metric determinant plays a role for the physics and,
in particular, for cosmology.
In both approaches, the allowed coordinate transformations are
restricted to those of unit Jacobian, so that the metric determinant
is a scalar under these restricted coordinate transformations.
In our approach, we then have that the metric determinant may enter 
certain terms of the matter Lagrange density.

The structure of the present explorative paper 
is somewhat different from that of the original
paper~\cite{Klinkhamer2022-ext-unimod}.
Here, we start from a hypothesis and an
action (Sec.~\ref{sec:Setup}), and then
investigate the resulting cosmology
(Secs.~\ref{sec:Cosmology-First-model}
and \ref{sec:Cosmology-Second-model}).
Having established an interesting cosmological behavior,
we turn towards one possible explanation of the hypothesis.
The idea is that
gravity may not be fundamental but is really an emerging
phenomenon from an underlying theory (Sec.~\ref{sec:Emergent-gravity}).
Concretely, we consider two possible realizations of emergent gravity.
The first realization (Sec.~\ref{subsec:Spacetime-crystal-Elasticity-tetrads})
relies on the
elasticity tetrads from a spacetime crystal~\cite{NissinenVolovik2019}
and has been elaborated in Ref.~\cite{Klinkhamer2022-ext-unimod}.
The second realization (Sec.~\ref{subsec:IIB-matrix-model}) 
is entirely new and uses
a nonperturbative formulation of superstring theory in the guise
of the IIB matrix model (references will be given later on).
That last model consists of $N \times N$ traceless Hermitian matrices, with
$10$ bosonic matrices and $8$ fermionic matrices. Somehow, these bosonic
matrices give rise to a classical spacetime and we will now argue
that appropriate perturbations of the relevant matrices can
give a nonstandard term in the matter Lagrange density
involving the metric determinant.
We present concluding remarks in Sec.~\ref{sec:Discussion}
and give technical details of the matrix-model calculation
in App.~\ref{app:IIB-matrix-model-calculation}.

\section{Setup}
\label{sec:Setup}

\subsection{Hypothesis}
\label{subsec:Hypothesis}

Our working hypothesis is that the field
$\sqrt{-g(x)}$, for $g(x) \equiv \det g_{\alpha\beta}(x)$,
corresponds to a physical quantity
(the spacetime metric $g_{\alpha\beta}$
has a Lorentzian signature, so that $g$ is negative).
Then, the only allowed coordinate transformations
$x^{\alpha} \to x'^{\,\alpha}$ are those of unit Jacobian,
\beq
\label{eq:Jacobian-unity}
\det\Big(\partial x'^{\,\alpha}/\partial x^{\beta}\Big) =1\,.
\eeq
In that case, it is possible that $\sqrt{-g(x)}$ also enters the matter
potential, as will be discussed in Sec.~\ref{subsec:Action}.

Incidentally, these restricted coordinate transformations
with \eqref{eq:Jacobian-unity} appear as well
in the unimodular-gravity approach to the cosmological constant
problem~\cite{Einstein1919,vanderBij-etal1982,Zee1983,BuchmuellerDragon1988,%
HenneauxTeitelboim1989} (a succinct review is given in
Sec.~VII of Ref.~\cite{Weinberg1989}).
The possibility of adding extra $\sqrt{-g}$ factors
in the matter action was already noted by Zee
on p. 220 of Ref.~\cite{Zee1983}, but was not pursued further.
Later, we will say more about the possible origin of
our extension of unimodular gravity,
but, at this moment, we just continue with the hypothesis.

\subsection{Action}
\label{subsec:Action}

We now investigate the implications of our hypothesis
by considering a relatively simple action, with
a standard real scalar field $\Chi(x)$ and
a single nonstandard term involving $\sqrt{-g(x)}$
in the matter Lagrange density.

The postulated action is given by~\cite{Klinkhamer2022-ext-unimod}
\bsubeqs\label{eq:action-S}
\beqa
S &=& S_{G}+ S_{M}^\text{\,(scalar)}
+ S_{M}^{\,(\Lambda-\text{plus})} + S_{N}\,,
\\[2.000mm]
\label{eq:action-G}
S_{G}&=& \int d^{4}x  \,\sqrt{-g}\;\frac{R}{16\pi G_{N}}\,,
\\[2.000mm]
\label{eq:action-SChi}
S_{M}^\text{\,(scalar)}&=& \int d^{4}x  \,\sqrt{-g}\;
\left[  \frac{1}{2}\,g^{\alpha\beta}\, \partial_{\alpha}\Chi\,\partial_{\beta}\Chi
+  \frac{1}{2}\,g_{2}\,M^{2}\,\Chi^{2}  \right]\,,
\\[2.000mm]
\label{eq:action-Lambdaplus}
S_{M}^{\,(\Lambda-\text{plus})}&=& \int d^{4}x  \,\sqrt{-g}\;
  \epsilon(\Lambda,\,n) \,,
\\[2.000mm]
\label{eq:action-SN}
S_{N} &=& - \mu\, \int d^{4}x  \; n(x)\,,
\\[2.000mm]
\label{eq:action-n-def}
n(x)&=& \sqrt{-g(x)} \;  M^{4} \geq 0 \,,
\eeqa
\esubeqs
where $g(x)$ is the determinant of the metric $g_{\alpha\beta}(x)$
with Lorentzian signature, $g_{2}$ a nonnegative constant,
and $1/M$  a fundamental length scale of the underlying theory
(recall that we are using natural units with $c=1$  and $\hbar=1$).
In \eqref{eq:action-Lambdaplus},
we simply take a linear dependence on $n$ for the potential,
\beqa\label{eq:epsilon-Ansatz}
\epsilon(\Lambda,\,n) &=& \Lambda  + \zeta\,n \,,
\eeqa
with a real parameter $\zeta>0$.
We emphasize that, strictly speaking,
the only new input is the single term $n \propto \sqrt{-g}$
in the potential \eqref{eq:epsilon-Ansatz},
which requires coordinate invariance to be restricted by
\eqref{eq:Jacobian-unity}. A possible
condensed-matter-type origin of the action \eqref{eq:action-S}
has been discussed in Ref.~\cite{Klinkhamer2022-ext-unimod}
and will be reviewed in Sec.~\ref{subsec:Spacetime-crystal-Elasticity-tetrads},
but this action can also have an entirely different origin.
In fact, a superstring-related origin will be discussed in
Sec.~\ref{subsec:IIB-matrix-model}.

In the resulting gravitational field equation,
\beq
\label{eq:Einstein-eq}
\frac{1}{8\pi G_{N}}
\left( R_{\alpha\beta}-\frac{1}{2}\,R\,g_{\alpha\beta}\right)=
\rho_\text{vac}\, g_{\alpha\beta}+T^{M}_{\alpha\beta}\,\big[\Chi\big]\,,
\eeq
we have, with the linear \textit{Ansatz} \eqref{eq:epsilon-Ansatz},
\bsubeqs\label{eq:rhovac-lambda-Ansaetze}
\beqa
\rho_\text{vac} &=&
\epsilon +n\,\frac{d \epsilon}{d n} -\mu\,M^{4}
= \Lambda  + 2\,\zeta\,n -\mu\,M^{4} \,,
\\[2mm]
\Lambda&=&  \lambda\,M^{4}\,,
\eeqa
\esubeqs
where the chemical potential $\mu \ne 0$ traces back to
the action term \eqref{eq:action-SN}
and $n$  has been defined by \eqref{eq:action-n-def}.

If we take
the covariant divergence of \eqref{eq:Einstein-eq}
and use the contracted Bianchi identities, we obtain the
following combined energy-momentum conservation relation:
\beq
\label{eq:combined-energy-momentumconservation}
\Big( \rho_\text{vac}\, g_{\alpha\beta}+T^{M}_{\alpha\beta}\Big)^{;\,\beta}
=0\,,
\eeq
where the semicolon stands for a covariant partial derivative.
If the matter component is separately conserved,
$\left( T^{M}_{\alpha\beta}\right)^{;\,\beta} =0$,
then equally so for the vacuum component,
$\left(  \rho_\text{vac}\, g_{\alpha\beta}\right)^{;\,\beta} =0$,
which implies
$\rho_\text{vac}^{\hspace*{4mm},\,\beta} =0$, where
the colon stands for a standard partial derivative.

Note that, in order to reach the Minkowski vacuum with
$\rho_\text{vac}=0$,
there is, for given chemical potential $\mu$
and the $\rho_\text{vac}$ expression \eqref{eq:rhovac-lambda-Ansaetze},
a restriction on the allowed cosmological constant,
\beq
\label{eq:lambda-restriction}
\lambda < \mu\,.
\eeq
Only for $(\mu-\lambda) > 0$,
is it possible to get $\rho_\text{vac}=0$ if
the positive vacuum variable $n$ adjusts itself to the value
\beq
\label{eq:n-Mink}
n_\text{Mink}
 =
M^{4}\,\frac{\,1}{2\,\zeta}\;\big(\mu- \lambda\big)\,.
\eeq

The restriction \eqref{eq:lambda-restriction}
can be evaded with a different
dependence on $n$ for the potential and a useful example is
\beqa\label{eq:epsilon-Ansatz-designer}
\widetilde{\epsilon}\,(\Lambda,\,n) =
\Lambda  + M^{-4}\, n^{2} +M^{12} \,n^{-2} \,.
\eeqa
The resulting gravitating vacuum energy density,
\beqa\label{eq:rhovac-Ansatz-designer}
\widetilde{\rho}_\text{vac} &=&
\widetilde{\epsilon} +n\,\frac{d\, \widetilde{\epsilon}}{d n} -\mu\,M^{4}
=
\Lambda  + 3\,M^{-4}\, n^{2} - M^{12} \,n^{-2} -\mu\,M^{4} \,,
\eeqa
can be nullified if $n$ takes the following unique positive value:
\beqa\label{eq:rhovac-Ansatz-designer-ntilde-Mink}
\widetilde{n}_\text{Mink}
 =
 M^{4}\,\sqrt{\frac{1}{6}\,
 \left[ \sqrt{12+(\mu-\lambda)^{2}} +(\mu-\lambda) \right]}  \,,
\eeqa
which is well defined for \emph{any} value of $(\mu-\lambda)$.
The vacuum energy density \eqref{eq:rhovac-Ansatz-designer}
will be used when we turn to cosmological solutions.

\section{Cosmology: First model}
\label{sec:Cosmology-First-model}

\subsection{Metric Ansatz}  
\label{subsec:Cosmology-Basic-model-Metric-Ansatz}

As the diffeomorphism invariance
of the model action \eqref{eq:action-S}
is restricted to transformations
of unit Jacobian, the appropriate spatially-flat
Robertson--Walker (RW) metric is given by~\cite{AlvarezFaedo2007}:
\beqa\label{eq:extRW-ds2}
\hspace*{-0mm}
ds^{2}
&=&
g_{\alpha\beta}(x)\, dx^{\alpha}\,dx^{\beta}
=
- \widetilde{A}(t)\;d t^{2}
+ \widetilde{R}^{\,2}(t)\;\delta_{m n}\,dx^{m}\,dx^{n}\,,
\eeqa
with the cosmic time coordinate $t$ from $x^{0}=c\,t=t$.
The spatial indices $m$, $n$ in \eqref{eq:extRW-ds2}
run over $\{1,\, 2,\, 3 \}$
and $\widetilde{R}(t)$ is the cosmic scale factor
[the tilde indicates the difference with the Ricci curvature
scalar appearing in \eqref{eq:action-G}].
Because the invariance transformations are restricted,
there is an additional \textit{Ansatz} function,
$\widetilde{A}(t)>0$.
We recover the standard spatially-flat RW metric
for $\widetilde{A}(t)=\text{const}>0$.
Remark that the extended RW metric \eqref{eq:extRW-ds2}
gives the vacuum variable
\beq\label{eq:n-for-ext-RW-metric}
n \propto \sqrt{-g} =(\widetilde{A}\,)^{1/2}\:|\widetilde{R}\,|^{3}\,,
\eeq
where the proportionality constant equals $M^4$ according to
\eqref{eq:action-n-def}. Having two \textit{Ansatz} functions
available, it is
possible to have constant $n$, also in an expanding universe. 

If, in the cosmological spacetime \eqref{eq:extRW-ds2},
the scalar field $\Chi$ is spatially homogeneous,
$\Chi=\Chi(t)$, then its energy-momentum tensor equals the one of
a perfect fluid having the following energy density
and pressure:%
\bsubeqs\label{eq:rho-P-wM-from-scalar}
\beqa
\rho_{\Chi}(t) &=&
\frac{1}{2}\,\frac{1}{\widetilde{A}(t)}\,\left(\frac{d\Chi(t)}{d t}\right)^{2}
+ \frac{1}{2}\,g_{2}\,M^{2}\,\Big(\Chi(t)\Big)^{2}\,,
\\[2mm]
P_{\Chi}(t) &=&
\frac{1}{2}\,\frac{1}{\widetilde{A}(t)}\,\left(\frac{d\Chi(t)}{d t}\right)^{2}
- \frac{1}{2}\,g_{2}\,M^{2}\,\Big(\Chi(t)\Big)^{2}\,.
\eeqa
If, moreover,  the scalar field $\Chi$ is rapidly oscillating,
$\Chi(t)=\Chi_{0}\,\cos(\omega\,t)$, then
the time-averages of the energy density and the pressure
give the following matter equation-of-state parameter:
\beqa
\label{eq:wM-from-scalar}
w_{M} &=&
\frac{\langle P_{\Chi}\rangle}{\langle \rho_{\Chi}  \rangle}
=
\frac{\omega^{2}/\widetilde{A} - g_{2}\,M^{2}}
{\omega^{2}/\widetilde{A} + g_{2}\,M^{2}}\,,
\eeqa
\esubeqs
under the assumption that the cosmological time scale relevant 
to $\widetilde{A}(t)$
is much larger than the oscillation periods $1/\omega$ or $1/M$.
Taking  $\omega^{2}/\widetilde{A}= 2\,g_{2}\,M^{2}$
in \eqref{eq:wM-from-scalar}, we obtain $w_{M} =1/3$.

In the following, we will work with this perfect fluid
instead of the original scalar $\Chi$ field
and take $w_{M} =1/3$, corresponding to
a gas of ultrarelativistic particles.

\subsection{Dimensionless ODEs}
\label{subsec:Cosmology-Basic-model-ODEs}

From now on, we set the model length scale $1/M$ equal to the
Planck length $1/E_\text{Planck}$,%
\beq
\label{eq:EPlanck-M-unity}
1/M = 1/E_\text{Planck} \equiv \sqrt{G_{N}}\,.
\eeq
We then introduce the following dimensionless quantities
(the chemical potential $\mu$ is already dimensionless):%
\bsubeqs\label{eq:dimensionless-variables}
\begin{align}
t &\to \tau\,,
\hspace*{-10mm}
&\rho_{\Chi}(t) &\to r_{\chi}(\tau)\,,
\hspace*{-10mm}
&\widetilde{\rho}_\text{vac}(t) &\to \widetilde{r}_\text{vac}(\tau)\,,
\\[2mm]
\Chi(t) &\to \chi(\tau)\,,
\hspace*{-10mm}
&P_{\Chi}(t) &\to p_{\chi}(\tau)\,,
\hspace*{-10mm}
&\widetilde{A}(t) &\to a(\tau)\,,
\\[2mm]
n(t) &\to n(\tau)\,,
\hspace*{-10mm}
&\Lambda &\to \lambda\,,
\hspace*{-10mm}
&\widetilde{R}(t)  &\to r(\tau)\,.
\end{align}
\esubeqs
where $n(\tau)$ is dimensionless and equal to
$\sqrt{-g(\tau)}=\sqrt{a(\tau)}\,|r(\tau)|^{3}$.
Also, we are using the vacuum energy density
from \eqref{eq:rhovac-Ansatz-designer}, which is marked
by a tilde.

From the field equations of the action \eqref{eq:action-S}
and with a homogeneous perfect fluid from the $\chi$ scalar,
we obtain the following
dimensionless ordinary differential equations (ODEs):
\bsubeqs\label{eq:dimensionless-ODEs}
\beqa
\label{eq:dimensionless-ODE-rchidoteq}
\hspace*{0mm}
&&\dot{r}_{\chi}
+  3\,(1+w_{M})\, \left(\frac{\dot{r}}{r}\right) \,r_{\chi}
 = 0 \,,
\\[2mm]
\label{eq:dimensionless-ODE-1stF}
\hspace*{0mm}
&&3\,\left( \frac{\dot{r}}{r} \right)^{2}
 =
8\,\pi \,a\, \Big(r_{\chi}+ \widetilde{r}_\text{vac} \Big)\,,
\\[2mm]
\label{eq:dimensionless-ODE-2ndF}
\hspace*{0mm}
&&\frac{2\,\ddot{r}}{r}
+\left( \frac{\dot{r}}{r} \right)^{2}
-\left( \frac{\dot{a}}{a} \right)\,\left( \frac{\dot{r}}{r} \right)
 =
- 8\,\pi \,a\,  \Big( w_{M}\,r_{\chi}-   \widetilde{r}_\text{vac} \Big) \,,
\\[2mm]
\label{eq:dimensionless-ODE-rvactilde}
\hspace*{0mm}
&& \widetilde{r}_\text{vac} =
\lambda   + 3\,a\,r^{6} -  a^{-1}\,r^{-6}  -\mu \,,
\eeqa
\esubeqs
where the overdot stands for
differentiation with respect to $\tau$.
These ODEs have two real parameters:
the matter equation-of-state parameter $w_{M} > -1$
and the combination $(\lambda-\mu$)
entering the vacuum energy density $\widetilde{r}_\text{vac}$.
Incidentally, the function $a(\tau)$ has been assumed to be positive.

It can be shown that the ODEs \eqref{eq:dimensionless-ODEs}
give the equation
\beq
\label{eq:dimensionless-ODE-rvacdot-zero}
\dot{r}_\text{vac}=0\,,
\eeq
so that the vacuum energy density stays constant over time.
This equation corresponds to
the energy-conservation equation of a homogeneous perfect
fluid with equation-of-state parameter $w_\text{vac}=-1$
[consider \eqref{eq:dimensionless-ODE-rchidoteq}
and replace $r_{\chi}$ by $\widetilde{r}_\text{vac}$
and $w_{M}$ by $w_\text{vac}=-1$].
In fact, \eqref{eq:dimensionless-ODE-rvacdot-zero} traces back
to  \eqref{eq:combined-energy-momentumconservation}
for matter with $\left( T^{M}_{\alpha\beta}\right)^{;\,\beta} =0$,
so that $\rho_\text{vac}^{\hspace*{4mm},\,\beta} =0$.  
In Sec.~\ref{sec:Cosmology-Second-model},
we will introduce a vacuum-matter energy exchange,
but here we just keep \eqref{eq:dimensionless-ODE-rvacdot-zero}
as it is.

\subsection{Analytic solutions}
\label{subsec:Analytic-solutions}

\subsubsection{Friedmann-type solution}
\label{subsubsec:Friedmann-type-solution}

We now present an exact Friedmann-type
solution of the ODEs \eqref{eq:dimensionless-ODEs}
for a general matter equation-of-state parameter $w_{M} > -1$.

Take the following \textit{Ansatz} functions for $\tau>0$:%
\bsubeqs\label{eq:functions-analytic-sol}
\beqa
a(\tau) &=& \alpha\;\tau^{-2\,p} \,,
\\[2mm]
r(\tau) &=& \alpha^{-1/6}\;\widehat{r}\; \tau^{p/3} \,,
\\[2mm]
r_{\chi}(\tau) &=& \alpha^{-1}\;\widehat{\chi}\;\tau^{-m}\,,
\eeqa
\esubeqs
with positive parameters $\alpha$, $p$, $\widehat{r}$, $\widehat{\chi}$,
and $m$. These \textit{Ansatz} functions have been designed to
produce a constant vacuum variable,
$\sqrt{-g}=\sqrt{a}\,|r|^{\,3} = \widehat{r}^{\;3}$.
The vanishing of $\widetilde{r}_\text{vac}$
from \eqref{eq:dimensionless-ODE-rvactilde} then gives
\beq
\label{eq:rhat-analytic-sol}
\widehat{r}_\text{\,sol} =
\left(\frac{1}{6}\,
 \left[ \sqrt{12+(\mu-\lambda)^{2}} +(\mu-\lambda) \right]\right)^{1/6}\,,
\eeq
where \eqref{eq:rhovac-Ansatz-designer-ntilde-Mink} has been used.

For the  \textit{Ansatz} functions \eqref{eq:functions-analytic-sol},
the dimensionless Ricci and Kretschmann curvature scalars read%
\bsubeqs\label{eq:R-K-analytic-sol}
\beqa
\label{eq:R-analytic-sol}
\mathcal{R}&=&
\frac{2}{3}\,
p\,\big(  5\,p -3 \big)\,\frac{1}{\alpha}\,{\tau}^{-2\,(1- p)}\,,
\\[2mm]
\label{eq:K-analytic-sol}
\mathcal{K}&=&
\frac{4}{27}\,p^{2}\,\big( 9 - 24\,p + 17\,p^{2}\big)\,
\frac{1}{\alpha^{2}}\,{\tau}^{-4\,(1- p)}\,.
\eeqa
\esubeqs
We now look for an expanding ($p>0$) Friedmann-type universe 
approaching Minkowski spacetime. These solutions
have a vanishing vacuum energy density throughout,
$\widetilde{r}_\text{vac}(\tau)=0$.

With the \textit{Ansatz} functions \eqref{eq:functions-analytic-sol},
the three ODEs \eqref{eq:dimensionless-ODEs}
for $\widehat{r} = \widehat{r}_\text{\,sol}$ from \eqref{eq:rhat-analytic-sol}
reduce to the following equations:
\bsubeqs\label{eq:ODES-functions-analytic-sol}
\beqa
\label{eq:ODES-rchieq-functions-analytic-sol}
0 &=&
\frac{1}{\alpha}\;
    \Big(  p\,\left(1 + w_{M}\right) - m \Big)\,
    \widehat{\chi}\,{\tau}^{-1 - m}  \,,
\\[2mm]
\label{eq:ODES-1stFeq-functions-analytic-sol}
0 &=&
\frac{p^{2}}{3\,{\tau}^{2}} - 8\,\pi\,\widehat{\chi} \,{\tau}^{-m - 2\,p}\,,
\\[2mm]
\label{eq:ODES-2ndFeq-functions-analytic-sol}
0 &=&
 \frac{p^{2}}{{\tau}^{2}} - \frac{2\,p}{3\,{\tau}^{2}} +
  8\,\pi\,w_{M}\,\widehat{\chi} \,{\tau}^{-m - 2\,p} \,.
\eeqa
\esubeqs
The exact solution of these equations has arbitrary $\alpha>0$ and
\bsubeqs\label{eq:constants-analytic-sol}
\beqa
p_\text{\,sol} &=& \frac{2}{3 + w_{M}}\,,
\\[2mm]
m_\text{\,sol} &=& \frac{2\,\left( 1 + w_{M} \right) }{3 + w_{M}}\,,
\\[2mm]
\widehat{\chi}_\text{\,sol} &=&
\frac{1}{6\,\pi \,{\left( 3 + w_{M} \right) }^{2}} \,,
\eeqa
\esubeqs
where $p_\text{\,sol}$ ranges over $(0,\,1)$
for $w_{M} \in (-1,\,+\infty)$.

The main points of this cosmology with $w_{M}=1/3$, for example,
are as follows:
\begin{itemize}
  \item[(i)]
an expanding Friedmann-type universe 
with cosmic scale factor
$r \sim \tau^{1/5}$.
  \item[(ii)]
a decreasing perfect-fluid energy density and pressure 
with $r_{\chi}(\tau)=3\,p_{\chi}(\tau) \sim \tau^{-4/5}$.
  \item[(iii)]
a cosmological constant $\lambda$ cancelled
by $\sqrt{-g}=\widehat{r}_\text{\,sol}$ from \eqref{eq:rhat-analytic-sol},
so that $r_\text{vac}(\tau)=0$.
  \item[(iv)]
the curvature scalars $\mathcal{R}(\tau) \sim 0$ and
$\mathcal{K}(\tau)\sim \tau^{-8/5}$,
approaching Minkowski spacetime.
\end{itemize}
Observe that, for a given value of $\mu$,   
we have not one solution but
a whole family of solutions, parametrized by the
value of the cosmological constant $\lambda$
which enters the solutions via \eqref{eq:rhat-analytic-sol}.

\subsubsection{De-Sitter-type solution}
\label{subsubsec:De-Sitter-type-solution}

In addition to an analytic Friedmann-type solution
with $\widetilde{r}_\text{vac}(\tau)=0$,
the ODEs \eqref{eq:dimensionless-ODEs}
can also have an analytic de-Sitter-type solution
with $\widetilde{r}_\text{vac}(\tau)=\text{const}>0$.

The \textit{Ansatz} functions for $\tau>0$
are taken as before, but now with a vanishing matter component,
\bsubeqs\label{eq:functions-analytic-deS-sol}
\beqa
a(\tau) &=& \alpha\;\tau^{-2\,p} \,,
\\[2mm]
r(\tau) &=& \alpha^{-1/6}\;\widehat{r}\; \tau^{p/3} \,,
\\[2mm]
r_{\chi}(\tau) &=& 0\,,
\eeqa
\esubeqs
for positive parameters $\alpha$, $p$, and $\widehat{r}$.
The general de-Sitter-type solution (denoted ``deS-gen-sol'') then
has the following parameters:%
\bsubeqs\label{eq:constants-analytic-deS-gen-sol}
\beqa
p_\text{deS-gen-sol} &=& 1\,,
\\[2mm]
\alpha_\text{deS-gen-sol} &=&
1/(24\,\pi\,\widetilde{r}_\text{vac-deS-gen-sol})\,,
\\[2mm]
\widetilde{r}_\text{vac-deS-gen-sol} &=&
\lambda + 3\,(\widehat{r}_\text{deS-gen-sol})^{6}
        - (\widehat{r}_\text{deS-gen-sol})^{-6}  -\mu
\\[2mm]
\widehat{r}_\text{deS-gen-sol} &>& 0\,.
\eeqa
\esubeqs
The corresponding dimensionless Ricci and Kretschmann curvature scalars read%
\bsubeqs\label{eq:R-K-analytic-deS-gen-sol}
\beqa
\label{eq:R-analytic-deS-gen-sol}
\mathcal{R}_\text{deS-gen-sol}&=&
\frac{4}{3}\,\frac{1}{\alpha_\text{deS-gen-sol}}\,,
\\[2mm]
\label{eq:K-analytic-deS-gen-sol}
\mathcal{K}_\text{deS-gen-sol}&=&
\frac{8}{27}\,\frac{1}{\alpha_\text{deS-gen-sol}^{2}}\,.
\eeqa
\esubeqs

The above solution has $\widehat{r}$ as a free parameter.
For $\lambda>0$, a special solution (denoted  ``deS-spec-sol'') has
vacuum energy density $\widetilde{r}_\text{vac}=\lambda$ if
the following parameters are chosen:
\bsubeqs\label{eq:constants-analytic-deS-spec-sol}
\beqa
p_\text{deS-spec-sol} &=& 1\,,
\\[2mm]
\alpha_\text{deS-spec-sol} &=&
 \frac{1}{24\,\pi\,\lambda}\,,
\\[2mm]
\widehat{r}_\text{deS-spec-sol} &=&
\left(\frac{1}{6}\,
 \left[ \sqrt{12+\mu^{2}} +\mu) \right]\right)^{1/6}\,.
\eeqa
\esubeqs
The corresponding dimensionless Ricci and Kretschmann curvature scalars
are given by \eqref{eq:R-K-analytic-deS-gen-sol}
with $\alpha_\text{deS-spec-sol}$ replacing $\alpha_\text{deS-gen-sol}$.

\section{Cosmology: Second model}
\label{sec:Cosmology-Second-model}

\subsection{Quantum-dissipative effects}

The cosmological model of Sec.~\ref{sec:Cosmology-First-model}
has a constant vacuum energy density $\widetilde{\rho}_\text{vac}$,
so that if $\widetilde{\rho}_\text{vac}$ is initially nonvanishing
it stays so later on.
Obviously, this conclusion can only change if there is a
mechanism to transfer vacuum energy to matter energy.

The authors of Ref.~\cite{KlinkhamerSavelainenVolovik2016} 
have discussed, in general terms, relaxation effects in $q$-theory.
A specific calculation~\cite{KlinkhamerVolovik-MPLA-2016},
for the standard spatially-flat Robertson--Walker metric
[i.e., $\widetilde{A}(t)=1$ in \eqref{eq:extRW-ds2}],
has considered particle production
by spacetime curvature~\cite{ZeldovichStarobinsky1977}.
The obtained Zeldovich--Starobinsky-type rate reads
\beq
\label{eq:Gamma-particle-production}
\Gamma_\text{particle-production}= \widehat{\gamma}\,
\left| \widetilde{R}^{-1} \; \frac{d \,\widetilde{R}}{d t} \right|\,R^{2}\,,
\eeq
where $\widehat{\gamma}$ is a calculated positive number,
$\widetilde{R}(t)$ is the cosmic scale function
of the metric \eqref{eq:extRW-ds2},
and $R(t)$ is the Ricci curvature scalar depending
on the two \textit{Ansatz} functions,
$R(t)=R\big[\widetilde{A}(t),\,\widetilde{R}(t)\big]$.

The energy of the produced particles must come from somewhere
and the obvious candidate is the vacuum. In that case,
the cosmic evolution of the vacuum  and matter energy densities
is given by
\bsubeqs\label{eq:VM-energy-exchange}
\beqa
\label{eq:VM-energy-exchange-a}
\frac{d \,\widetilde{\rho}_\text{vac}}{d t} + \cdots
&=&
- \Gamma_\text{particle-production}\,,
\\[1mm]
\label{eq:VM-energy-exchange-b}
\frac{d \rho_M}{d t} + \cdots
&=&
+ \Gamma_\text{particle-production}\,,
\eeqa
\esubeqs
because of energy conservation \eqref{eq:combined-energy-momentumconservation}.
The equations \eqref{eq:VM-energy-exchange-a}
and \eqref{eq:VM-energy-exchange-b} are manifestly
time-reversal noninvariant for
the source term from \eqref{eq:Gamma-particle-production}.
This time-reversal noninvariance is to be
expected for a dissipative effect, in fact, a quantum-dissipative effect
as particle creation or annihilation is a true quantum phenomenon.

\subsection{ODEs with vacuum-matter energy exchange}
\label{subsec:ODEs-with-VM-energy-exchange}

We now consider a relativistic matter component with a constant
equation-of-state parameter $w_{M} \equiv P_{\chi}/\rho_{\chi} =1/3$
and add a positive source term $\Gamma$ on the right-hand side of
\eqref{eq:dimensionless-ODE-rchidoteq}.
We then need to determine how this addition feeds into the other
two ODEs, \eqref{eq:dimensionless-ODE-1stF}
and \eqref{eq:dimensionless-ODE-2ndF}.
We switch to the dimensionless variables \eqref{eq:dimensionless-variables}
and take three steps.

In step 1, we add, as mentioned above, 
a source term $\Gamma$ to the right-hand side
of \eqref{eq:dimensionless-ODE-rchidoteq} for $w_{M}=1/3$ to get
\bsubeqs\label{eq:dimensionless-modified-ODEs}
\beqa
\label{eq:dimensionless-modified-ODEs-eq1}
\hspace*{0mm}
&&
\dot{r}_{\chi} + 4\,
\left(\frac{\dot{r}}{r}\right)\,r_{\chi} = \Gamma \,,
 \eeqa
 where $\Gamma$ still needs to be specified.

In step 2, we eliminate $r_{\chi}$
by taking the sum of one third of
\eqref{eq:dimensionless-ODE-1stF}
and \eqref{eq:dimensionless-ODE-2ndF} for $w_{M}=1/3$,
\beqa
\label{eq:dimensionless-modified-ODEs-eq2}
&&
\frac{1}{8\,\pi \,a}\,
\left[
\frac{2\,\ddot{r}}{r} +2\,
\left( \frac{\dot{r}}{r} \right)^{2}
-\left( \frac{\dot{r}}{r} \right)\,\left( \frac{\dot{a}}{a} \right)
\right] =
\frac{4}{3}\,
\widetilde{r}_\text{vac}\,.
\eeqa
We observe that the left-hand side
of \eqref{eq:dimensionless-modified-ODEs-eq2}
is proportional to the Ricci scalar,
so that $\mathcal{R} \propto \widetilde{r}_\text{vac}$.
This observation will be used later on.

In step 3,
we take the derivative of \eqref{eq:dimensionless-ODE-1stF},
use \eqref{eq:dimensionless-modified-ODEs-eq1} to eliminate $\dot{r}_{\chi}$,
use \eqref{eq:dimensionless-ODE-1stF} to eliminate $r_{\chi}$,
use the $\ddot{r}$ expression from \eqref{eq:dimensionless-modified-ODEs-eq2},
and get
\beqa
\label{eq:dimensionless-modified-ODEs-eq3}
\hspace*{0mm}
&&
 \dot{\widetilde{r}}_\text{vac} = -\Gamma\,,
\\[2.0mm]
\label{eq:dimensionless-modified-ODEs-rvactilde}
\hspace*{0mm}
&&
\widetilde{r}_\text{vac} =\lambda   + 3\,a\,r^{6} -  a^{-1}\,r^{-6}  -\mu  \,,
\eeqa
\esubeqs
where the explicit $\widetilde{r}_\text{vac}$ expression has  been
recalled in the last equation.
For completeness, we give the original first-order Friedman equation,
\beqa
\label{eq:dimensionless-modified-ODEs-1stFeq}
\hspace*{0mm}
&&
3\,\left( \frac{\dot{r}}{r} \right)^{2} =
8\,\pi \,a\, \Big(r_{\chi}+\widetilde{r}_\text{vac} \Big)\,,
\eeqa
which, if it holds initially for the
solution of the ODEs \eqref{eq:dimensionless-modified-ODEs},
will be satisfied at subsequent times (this will make for a valuable
diagnostic of the numerical accuracy later on).

We remark that,
for Minkowski spacetime with $a(\tau)=r(\tau)=1$
in the dimensionless version of \eqref{eq:extRW-ds2},
we have $\dot{r}_{\chi} = \Gamma$
from \eqref{eq:dimensionless-modified-ODEs-eq1}
and $\dot{\widetilde{r}}_\text{vac} = - \Gamma$
from \eqref{eq:dimensionless-modified-ODEs-eq3},
showing the direct vacuum-matter energy exchange
provided $\Gamma$ is nonvanishing.

The next point is to simplify the expression for $\Gamma$
so that the numerics runs efficiently. We take
\bsubeqs\label{eq:Gamma-gammatilde}
\beqa
\label{eq:Gamma}
\Gamma(\tau) &=& \widetilde{\gamma}(\tau)\;
\big|\dot{r}(\tau)/r(\tau)\big|\;\,
\Big( \widetilde{r}_\text{vac}(\tau) \Big)^{2}\,,
\\[2.0mm]
\label{eq:gammatilde}
\widetilde{\gamma}(\tau) &=& \gamma\;
\left[
\frac{\tau^{2}-\tau_\text{bcs}^{2}}{\tau^{2}+1}
\right]^{2}\,,
\\[2.0mm]
\gamma &\geq& 0\,,
\eeqa
\esubeqs
for initial boundary conditions at  $\tau=\tau_\text{bcs}$
and a nonnegative constant $\gamma$.
The expression \eqref{eq:Gamma} basically has the structure
of \eqref{eq:Gamma-particle-production},
because the left-hand side
of \eqref{eq:dimensionless-modified-ODEs-eq2}  is proportional
to the Ricci scalar, so that $\mathcal{R} \propto \widetilde{r}_\text{vac}$.
We have also added a smooth switch-on function
$\widetilde{\gamma}(\tau)$ in \eqref{eq:Gamma-gammatilde},
in order to improve the
numerical evaluation of the ODEs.

Observe, again, that the
ODEs \eqref{eq:dimensionless-modified-ODEs-eq1}
and \eqref{eq:dimensionless-modified-ODEs-eq3}
with source term \eqref{eq:Gamma-gammatilde} are time-reversal noninvariant.
The basic structure of the resulting vacuum-energy equation,
\beq
\label{eq:dimensionless-modified-ODEs-rvacdoteq-structure}
 \dot{\widetilde{r}}_\text{vac}
 = - \widetilde{\gamma}\;
\big|\dot{r}/r\big|\;\big( \widetilde{r}_\text{vac} \big)^{2}\,,
\eeq
is similar to the one discussed in
Refs.~\cite{KlinkhamerVolovik-MPLA-2016,Klinkhamer2022-ext-unimod},
where an analytic solution for the vacuum energy density
was obtained and where that solution was found
to drop to zero as $\tau\to\infty$.

The exact Friedmann-type
solution of Sec.~\ref{subsec:Analytic-solutions}
carries over to
the modified ODEs \eqref{eq:dimensionless-modified-ODEs}
with source term \eqref{eq:Gamma-gammatilde} .
The reason is simply that this source term $\Gamma$ vanishes if
$\widetilde{r}_\text{vac}=0$, which is precisely the case for
our Friedmann-type solution.

\subsection{Numerical results}
\label{subsec:Numerical-results-modified-ODES}

Extensive numerical results were reported in
Ref.~\cite{Klinkhamer2022-ext-unimod},
establishing, in particular, the attractor behavior towards
Minkowski spacetime. These numerical results were based on the
linear \textit{Ansatz} \eqref{eq:epsilon-Ansatz}.
Here, we give some complementary numerical results
based on the extended \textit{Ansatz} \eqref{eq:epsilon-Ansatz-designer},
which confirm the previously found attractor behavior.

\begin{figure}[t]
\vspace*{0mm}
\begin{center}
\hspace*{0mm}
\includegraphics[width=1\textwidth]{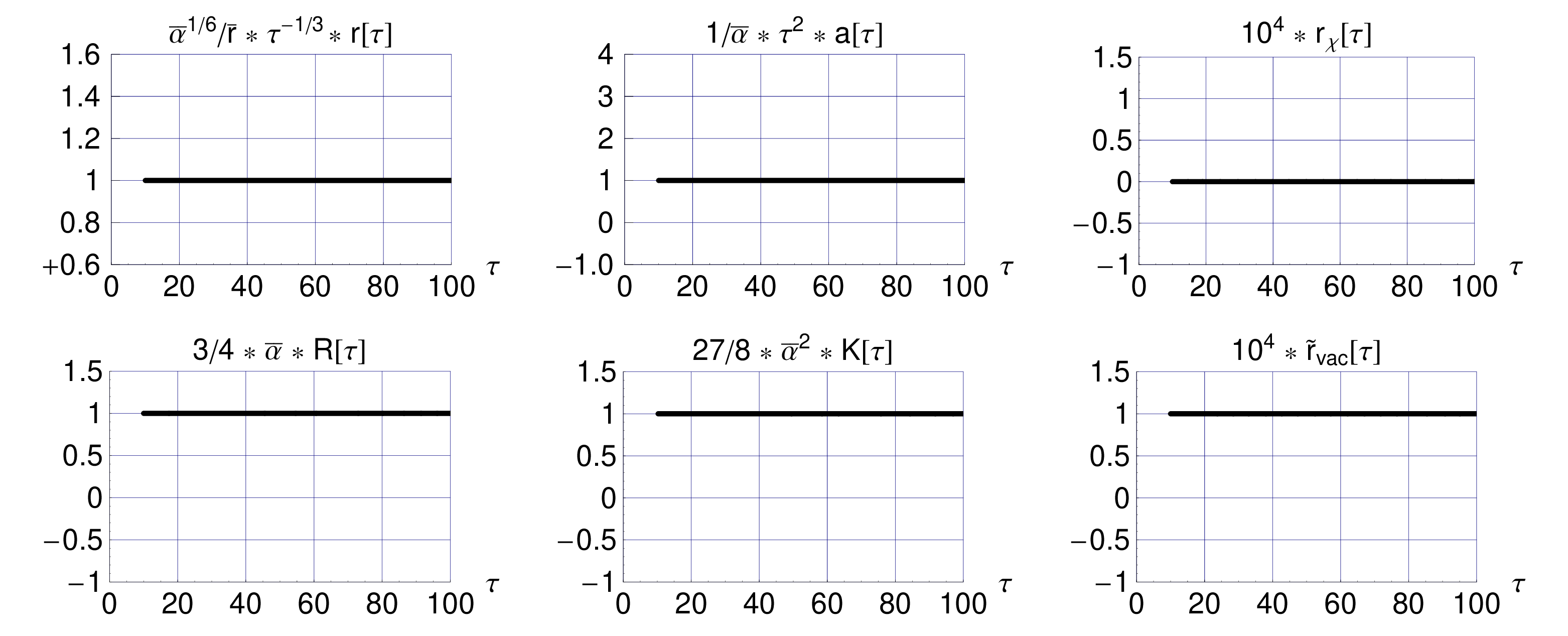}
\end{center}
\caption{Numerical solution of the ODEs
\eqref{eq:dimensionless-modified-ODEs} with
source term \eqref{eq:Gamma-gammatilde} and parameters $w_{M}=1/3$,
\mbox{$\mu=-3$,}   
$\lambda=10^{-4}$, and $\gamma=0$
(quantum-dissipative effects inoperative).
The initial boundary conditions are
taken from the analytic de-Sitter-type
solution \eqref{eq:functions-analytic-deS-sol}
and \eqref{eq:constants-analytic-deS-spec-sol},
with $\overline{\alpha}\equiv
\alpha_\text{deS-spec-sol}=132.629$
and $\overline{r} \equiv r_\text{deS-spec-sol}=0.800821$. 
Specifically, the boundary conditions at $\tau=\tau_\text{bcs}=10$ are:
$\{ a,r,\dot{r},r_{\chi}\}$ $=$
$\{  1.32629119, 0.764004165, 0.02546680549, 0 \}$,  
where the $\dot{r}$ value has been obtained from
the first Friedman equation \eqref{eq:dimensionless-modified-ODEs-1stFeq}.
The top row shows the three basic variables: the two metric functions
$r(\tau)$ and $a(\tau)$, and the dimensionless matter energy density $r_{\chi}$.
The bottom row shows three derived quantities:
the dimensionless Ricci curvature scalar $\mathcal{R}$,
the dimensionless Kretschmann curvature scalar $\mathcal{K}$,
and the dimensionless gravitating vacuum energy density
$\widetilde{r}_\text{vac}$ from \eqref{eq:dimensionless-modified-ODEs-rvactilde}.
The vacuum energy density from the initial conditions
is $\widetilde{r}_\text{vac}(\tau_\text{bcs})=1 \times 10^{-4}$,
which stays essentially constant.
}
\label{fig:num-sol-lambda0pt0001-deSbcs-gamma0-modODEs}
\end{figure}

\begin{figure}[t]
\begin{center}
\hspace*{0mm}
\includegraphics[width=1\textwidth]{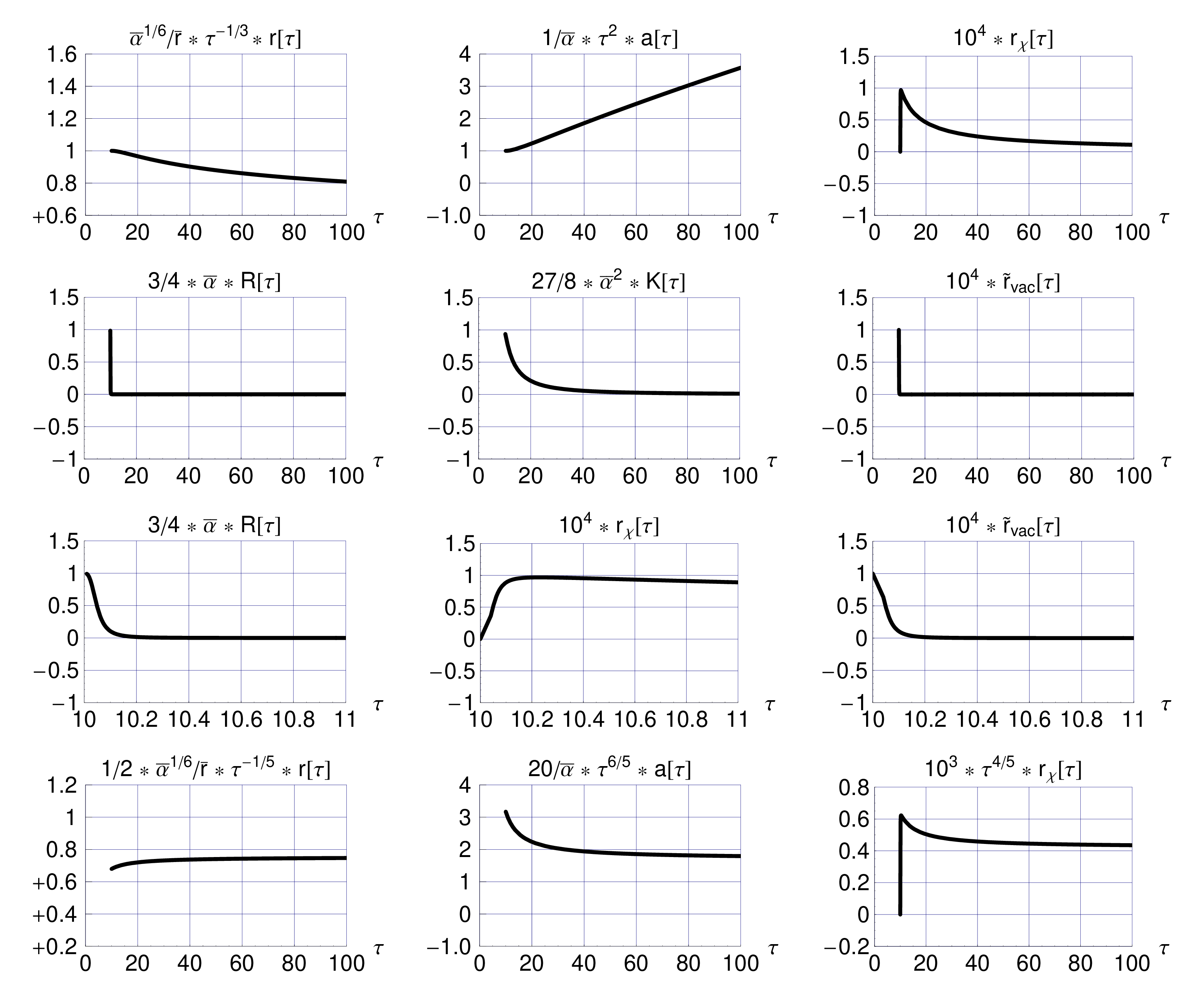}
\end{center}
\caption{Numerical solution of the ODEs
\eqref{eq:dimensionless-modified-ODEs} with
source term \eqref{eq:Gamma-gammatilde}, where
the boundary conditions at $\tau=\tau_\text{bcs}=10$
and the model parameters are the same as in
Fig.~\ref{fig:num-sol-lambda0pt0001-deSbcs-gamma0-modODEs},
but now with $\gamma=2 \times 10^{11}$
(quantum-dissipative effects operative).
The vacuum energy density
is  initially $\widetilde{r}_\text{vac}(10)=1 \times 10^{-4}$
and drops to $\widetilde{r}_\text{vac}(100)\sim 4 \times 10^{-10}$.
The third row shows the smooth behavior near the initial boundary conditions
and the fourth row the asymptotic Friedmann-type behavior.
}
\label{fig:num-sol-lambda0pt0001-deSbcs-gamma2E11-modODEs}
\vspace*{0mm}
\end{figure}

We start from the special de-Sitter-type configuration as given
in Sec.~\ref{subsubsec:De-Sitter-type-solution}.
Numerical results, for $\mu=-3$ and $\lambda=10^{-4}$, are presented
in Figs.~\ref{fig:num-sol-lambda0pt0001-deSbcs-gamma0-modODEs}
and \ref{fig:num-sol-lambda0pt0001-deSbcs-gamma2E11-modODEs}
with two values of the vacuum-matter-energy-exchange coupling constant $\gamma$.
The numerical solution of Fig.~\ref{fig:num-sol-lambda0pt0001-deSbcs-gamma0-modODEs}
with $\gamma=0$ essentially reproduces the special de-Sitter-type solution of
Sec.~\ref{subsubsec:De-Sitter-type-solution}, whereas
the numerical solution
of Fig.~\ref{fig:num-sol-lambda0pt0001-deSbcs-gamma2E11-modODEs}
with $\gamma=2 \times 10^{11}$ shows the rapid reduction
of the vacuum energy density $\widetilde{r}_\text{vac}$
and the approach to the analytic Friedmann-type solution
of Sec.~\ref{subsubsec:Friedmann-type-solution}.
The results in Figs.~\ref{fig:num-sol-lambda0pt0001-deSbcs-gamma0-modODEs}
and \ref{fig:num-sol-lambda0pt0001-deSbcs-gamma2E11-modODEs}
resemble those in Figs.~9 and 10 of Ref.~\cite{Klinkhamer2022-ext-unimod},
but there are significant differences as regards
the value of the chemical potential $\mu$,
the initial condition on $r(\tau)$,
and the value of the scaling factor $\overline{r}$.

We have two important remarks regarding the comparison of the
vacuum energy density results
obtained here and those obtained previously.
First, we note that $(\mu-\lambda) < 0$ does not allow for
the nullification of the vacuum energy density for the case 
of the linear $\epsilon$ \textit{Ansatz} \eqref{eq:epsilon-Ansatz},
which was the \textit{Ansatz} used in Ref.~\cite{Klinkhamer2022-ext-unimod}.
Second, the $r_\text{vac}$ panels of Fig.~10
in Ref.~\cite{Klinkhamer2022-ext-unimod}
and the $\widetilde{r}_\text{vac}$ panels of
Fig.~\ref{fig:num-sol-lambda0pt0001-deSbcs-gamma2E11-modODEs} here
are identical within the numerical accuracy, because the resulting
ODE \eqref{eq:dimensionless-modified-ODEs-rvacdoteq-structure}
for $\widetilde{r}_\text{vac}(\tau)$
and the corresponding ODE for $r_\text{vac}(\tau)$
in Ref.~\cite{Klinkhamer2022-ext-unimod} have the same structure
and the same boundary value $10^{-4}$ at $\tau=10$,
the only difference being the
``internal'' structure of $\widetilde{r}_\text{vac}(\tau)$
and $r_\text{vac}(\tau)$.

\newpage  
To summarize, we have shown in Ref.~\cite{Klinkhamer2022-ext-unimod}
and the present paper that, in principle,
the cosmological constant $\Lambda$ 
can be cancelled by the field $\sqrt{-g}$  
and appropriate quantum-dissipative effects
(in principle, there can be other vacuum-matter 
energy-exchange mechanisms). For completeness,
we have also given, in App.~C of Ref.~\cite{Klinkhamer2022-ext-unimod},
numerical results on the readjustment after a phase transition.

\newpage  
\section{Emergent gravity}
\label{sec:Emergent-gravity}

\subsection{Spacetime crystal: Elasticity tetrads}
\label{subsec:Spacetime-crystal-Elasticity-tetrads}

A possible explanation of the hypothesis presented
in Sec.~\ref{subsec:Hypothesis} is that gravity
is not a fundamental interaction but rather an emergent phenomenon.
One explicit suggestion was outlined in
Ref.~\cite{Klinkhamer2022-ext-unimod} and will
be briefly reviewed here. Another explicit suggestion
will be discussed in the next subsection.

In Sec.~IV of Ref.~\cite{Klinkhamer2022-ext-unimod},
we have considered the following matter Lagrange density term
for a real scalar field $\phi$:%
\bsubeqs\label{eq:example-epsilon-n-epsilontilde}
\beqa\label{eq:example-epsilon}
\epsilon(\phi,\,n) &=&
\widetilde{\epsilon}\,(\phi)\left(1+ \zeta\, M^{-4}\,n \right)\,,
\eeqa
with a positive dimensionless constant $\zeta$,
the definition
\beqa
\label{eq:example-n}
n(x)\equiv E(x)=M^{4}\,\sqrt{-g(x)}\,,
\eeqa
and now an explicit example of the function $\widetilde{\epsilon}$,
\beqa
\label{eq:example-epsilontilde}
\widetilde{\epsilon}\,(\phi) &=& M^{4}\,+ \frac{1}{2}\,M^{2}\,\phi^{2}\,,
\eeqa
\esubeqs
where $M$ is a single mass scale (possibly of the order of the Planck energy).
The resulting gravitating vacuum energy density
is~\cite{Klinkhamer2022-ext-unimod}
\beqa
\label{eq:example-rhoV}
\rho_\text{vac}(\phi,n) &=&
\widetilde{\epsilon}\,(\Phi)
\left(1+ 2 \,\zeta\,M^{-4}\,n\right)- \mu \,M^{4}\,,
\eeqa
where $\mu$ is the chemical potential corresponding to
the conservation of the spacetime points of
a hypothetical crystal.

The quantity $E$ in \eqref{eq:example-n}
stands for the determinant of the elasticity
tetrads of the spacetime crystal
(see Sec.~II of Ref.~\cite{Klinkhamer2022-ext-unimod}
and especially Ref.~\cite{NissinenVolovik2019} for further background
on elasticity tetrads).
With the assumption of gravity arising from these
elasticity tetrads, $E$ is then identified with
the square root of minus the metric determinant, $E\propto\sqrt{-g}$,
where the minus sign holds for a Lorentzian signature of the
emergent spacetime metric $g_{\alpha\beta}$.

Even though the elasticity tetrads can, in principle, produce
a nonstandard term $\half\,M^{2}\,\phi^{2}\,\sqrt{-g}$
as in \eqref{eq:example-epsilon}, it is
not clear how this would really come about. In this respect,
an explicit suggestion based on a matrix model
is perhaps more compelling,
and such a suggestion will be discussed next.

\subsection{IIB matrix model}
\label{subsec:IIB-matrix-model}

\subsubsection{Emerging spacetime}
\label{subsubsec:Emerging-spacetime}

It has been submitted that the IIB matrix  
model~\cite{IKKT-1997,Aoki-etal-review-1999}
can give rise to some type of spacetime lattice
and an emergent spacetime metric.
The authors of Ref.~\cite{Aoki-etal-review-1999}, in particular,
have argued that
``the space-time is dynamically determined from
the eigenvalue distributions of the matrices'' (quote from the Abstract)
and that
``the invariance under a permutation of the
eigenvalues leads to the invariance of the
low-energy effective action under general
coordinate transformations'' (quote from Sec.~4.2).
Most likely, the basic idea is correct,
but, strictly speaking, the matrices
in Refs.~\cite{IKKT-1997,Aoki-etal-review-1999}
are mere integration variables
and there is no small dimensionless parameter
to motivate a saddle-point approximation.
A possible solution of this puzzle has been
provided by the recent suggestion that
the so-called master field controls the
emergence of a classical spacetime,
indeed as eigenvalues but now the
eigenvalues of the IIB-matrix-model master-field matrices.

This new conceptual idea was proposed in
Ref.~\cite{Klinkhamer2021-master}, which also
contains an explicit procedure
on how to extract the classical spacetime
from the IIB-matrix-model master-field matrices.
Meanwhile, several follow-up research papers have appeared in
Ref.~\cite{Klinkhamer2021-IIBmm-regulBB,%
Klinkhamer2021-IIBmm-master-field-sol-1,%
Klinkhamer2021-IIBmm-master-field-sol-2,%
Klinkhamer2021-IIBmm-master-field-sol-3}, together with
two comprehensive
reviews~\cite{Klinkhamer2021-EpiphanyConf,Klinkhamer2022-CorfuConf}.

Here, we intend to show that this matrix-model approach can also provide
a possible explanation of nonstandard terms in the
matter Lagrange density involving the metric determinant.
As a preparation for this discussion, we have collected
in App.~\ref{subapp:Emerging-spacetime-points} 
the main steps of how the information about the
emergent spacetime points $\widehat{x}^{\,\alpha}_{k}$
is encoded in the master-field matrices.
These discrete spacetime points $\widehat{x}^{\,\alpha}_{k}$
are labeled by $k \in \{1,\,\ldots ,\,  K\}$,
with $K$ a divisor of $N$.

\subsubsection{Discrete effective action}
\label{subsubsec:Discrete-effective-action}

From appropriate perturbations of the
master-field matrices $\widehat{A}^{\,\alpha}$
(restricting to $D=4$ ``large'' Euclidean dimensions), the following
effective action can be obtained for a low-energy scalar
degree of freedom $\phi$  propagating over the discrete
spacetime points $\widehat{x}^{\,\alpha}_{k}$:   
\beqa
\label{eq:Seff-phi-discrete}
S_\text{eff}[\phi_{k},\,\eta_{k},\,\ldots] &\supset&
\sum_{k=1}^{K}\;\sum_{l=1}^{K}\; \frac{1}{2}\,
\widetilde{f}\big(\widehat{x}_{k}-\widehat{x}_{l}\big)\,
\ell^{2}\,\big( \phi_{k}- \phi_{l}  \big)^{2}
+ \sum_{k=1}^{K}\;\frac{1}{2}\,\widetilde{m}^{2}\,\ell^{2}\, \big( \phi_{k}\big)^{2}
\nonumber\\
&&
+\sum_{k=1}^{K}\;\sum_{l=1}^{K}\;
\widetilde{h}\big(\widehat{x}_{k}-\widehat{x}_{l}\big)\,
\big( \phi_{k}\big)^{2}\, \big( \eta_{l}\big)^{2},
\eeqa
where $\widetilde{f}(x)$ and $\widetilde{h}(x)$
are steep dimensionless functions centered on $x=0$,
the matrix perturbations $\phi_{k}$ and $\eta_{k}$ have the dimension of length,
$\widetilde{m}$ is dimensionless, and $\ell$ is the model length
scale~\cite{Aoki-etal-review-1999,Klinkhamer2021-master}
(on the length scale issue, see also the last paragraph of Sec.~5 in
Ref.~\cite{Klinkhamer2022-CorfuConf}).
The effective action can, of course, be expected 
to contain fields
of all spins, but considering only scalar fields
suffices for our purpose.
Details on how the discrete scalar-field effective 
action \eqref{eq:Seff-phi-discrete}
can be obtained are given in App.~\ref{subapp:Perturbations}.  

We observe the permutation symmetry of the
result \eqref{eq:Seff-phi-discrete}:
\bsubeqs\label{eq:permutation-symmetry}
\beqa
\widehat{x}^{\,\alpha}_{k} &\rightarrow& \widehat{x}^{\,\alpha}_{\sigma(k)}\,,
\quad
\sigma  \in S_{K}\,,
\eeqa
\esubeqs
with corresponding changes of the matrix perturbations,
$\phi_{k} \rightarrow \phi_{\sigma(k)}$
and $\eta_{k} \rightarrow \eta_{\sigma(k)}$.
The role of this permutation symmetry
will be discussed further in Sec.~\ref{subsubsec:Nonstandard-action-term}.

\subsubsection{Standard action terms in the continuum}
\label{subsubsec:Standard-action-terms}

The first two terms in \eqref{eq:Seff-phi-discrete}
were discussed in
Refs.~\cite{Aoki-etal-review-1999,Klinkhamer2021-master},
but the last term is new.
These first two terms give the following continuum effective action
for a real scalar field $\phi(x)$ of mass dimension 1:
\beq
\label{eq:Seff-kin-mass}
S_\text{eff}^\text{kin+mass}[\phi(x)] \sim
\int d^4x \,\sqrt{|g(x)|}\;
\left[\frac{1}{2}\;g^{\alpha\beta}(x)\;
\partial_{\alpha}\phi(x)\,\partial_{\beta}\phi(x)\,
+ \frac{1}{2}\,m^{2}\;\phi^{2}\right]\,,
\eeq
in terms of an emergent inverse metric $g^{\alpha\beta}$
and a classical dilaton field $\Phi$ 
(here, this dilaton field has been assumed constant 
and has been normalized away).  
Specifically, we have for these emerging fields
\bsubeqs\label{eq:emergent-inverse-metric-root-det}
\beqa \label{eq:emergent-inverse-metric}
g^{\alpha\beta}(x) &\sim&
\int_{\mathbb{R}^{4}} d^{4}y\;
\rho(y)\; \ell^{-2}\,
(x-y)^{\,\alpha}\,(x-y)^{\beta}\;r(x,\,y)\;f(x-y)\,,
\\[2mm]
\label{eq:emergent-root-det}
\sqrt{|g(x)|}
&\propto&
\rho(x)\,,
\eeqa
\esubeqs
with $g \equiv \det g_{\alpha\beta}$.
The square root $\sqrt{|g(x)|}$ can also be written as
$\sqrt{\pm g(x)}$, where the $\pm$ signs refer to the
Euclidean or Lorentzian signatures of the emerging spacetime metric; 
see below for further comments. The quantities
$\rho(x)$ and $r(x,\,y)$ entering the expressions
\eqref{eq:emergent-inverse-metric-root-det} result from
the distributions and correlations of the
spacetime points $\widehat{x}^{\,\alpha}_{k}$
obtained from the master field; their definitions are given 
in App.~\ref{subapp:Emerging-spacetime-points}.  
The quantity $f(x)$ entering the expression
\eqref{eq:emergent-inverse-metric} traces back
to the first term in \eqref{eq:Seff-phi-discrete},
which results from perturbations of the master-field matrices;
see App.~\ref{subapp:Perturbations}.

Equations \eqref{eq:emergent-inverse-metric}
and \eqref{eq:emergent-root-det}
have essentially been given as Eqs.~(4.17) and (4.18) in
Ref.~\cite{Aoki-etal-review-1999}, but here we have
made clear precisely which matrices are considered
for the eigenvalues,
namely the master-field matrices $\widehat{A}^{\,\alpha}$
(the heuristics of the spacetime extraction from the master-field matrices
is explained in Sec.~4.4 of Ref.~\cite{Klinkhamer2021-EpiphanyConf}).
For completeness, we mention that there have also been 
other approaches on getting an effective metric field    
from the IIB matrix model (see, e.g.,
Refs.~\cite{Sakai2019,Steinacker2022,Brahma-etal2022} and
references therein). 

At this moment, we have three technical remarks,
which can be skipped in a first reading.
The first technical remark is that
the signature of the emerging spacetime metric depends
on the structure of the correlation functions of the
spacetime points. Toy-model calculations have been
presented in App.~D of Ref.~\cite{Klinkhamer2021-EpiphanyConf},
which show that certain deformation parameters in
the correlation functions allow for a continuous change
from a Euclidean to a Lorentzian signature
(passing through a degenerate metric with a vanishing eigenvalue).

The second technical remark is that,
assuming the matrix size $N=K\,n$ to be large enough and
the block size $n$ to be of the order of the band 
width $\Delta N$ 
(see App.~\ref{subapp:Emerging-spacetime-points} for details),
we need not average the $\rho$
functions appearing on the right-hand sides of
\eqref{eq:emergent-inverse-metric-root-det}.
This averaging would be over different block sizes ($n$)
and over different block positions along the diagonals
of the master-field matrices . 
But these explicit averages would not be necessary
if we have a genuine master field at an effectively infinite $N$
(for a  general discussion of the role of master fields, 
see, in particular, Ref.~\cite{Carlson-etal-1983}).

The third technical remark is that the fermion dynamics
plays an important role for the bosonic master-field
matrices $\widehat{A}^{\,\alpha}$, as they are the solution
of the so-called master-field equation which has a Pfaffian
term due to the fermions. The master-field equation and its
solutions have been studied in three recent
papers~\cite{Klinkhamer2021-IIBmm-master-field-sol-1,%
Klinkhamer2021-IIBmm-master-field-sol-2,%
Klinkhamer2021-IIBmm-master-field-sol-3} and have been
reviewed in Ref.~\cite{Klinkhamer2022-CorfuConf}.
(Recall that the fermion dynamics also played an important
role in the original discussion of Ref.~\cite{Aoki-etal-review-1999}
by providing the so-called Boltzmann weights in the
graphs considered.)

\subsubsection{Nonstandard action term in the continuum}
\label{subsubsec:Nonstandard-action-term}

Now, turn to the third term in \eqref{eq:Seff-phi-discrete},
which is new and has been ``derived'' 
in App.~\ref{subapp:Perturbations}.   
For a steep function $\widetilde{h}\big(\widehat{x}_{k}-\widehat{x}_{l}\big)$,
having $\widetilde{h}\sim 0$ for $\widehat{x}_{k}\ne \widehat{x}_{l}$,
and constant perturbations $\eta_{k}$,
\beq
\label{eq:constant-eta-k}
 \eta_{k} = \overline{\eta}\,,
\eeq
we get the following nonstandard term in the continuum effective action:
\beq
\label{eq:Seff-nonstandard}
S_\text{eff}^\text{(nonstandard)}[\phi(x)] \sim
\int d^4x \,\sqrt{|g(x)|}\;
\Big[\;\phi^{2}(x)\;\sqrt{|g(x)|}\;\eta^{2}\Big]\,,
\eeq
where $\eta$ is a constant real scalar field of mass dimension 1 tracing back 
to the rescaled constant matrix perturbation $\overline{\eta}/\ell^{2}$.

Incidentally, by taking also $\phi_{k}$ constant,
$\phi_{k} = \overline{\eta}$,
we get the action term \eqref{eq:Seff-nonstandard}
with the integrand $\big[\eta^{2}\,\sqrt{|g(x)|}\:\eta^{2}\big]$,
corresponding to the linear $\zeta$ term appearing in our previous
$\epsilon$ \textit{Ansatz} \eqref{eq:epsilon-Ansatz}.
It appears impossible to get, in this way, negative powers of
$n \propto \sqrt{|g|}$, as used in
the extended $\epsilon$ \textit{Ansatz} \eqref{eq:epsilon-Ansatz-designer}.
Still, there may appear divergent powers series in $n$, 
which, for example, sum to $\epsilon$ functions of the form 
$\lambda +n^2/(n-1)$ or $\lambda +n \, \cos(2\pi\,n)$,
allowing for the cancellation of any value of
$(\lambda-\mu)$ in the corresponding $r_\text{vac}$ functions.

The action term \eqref{eq:Seff-nonstandard} is, of course, only invariant
under restricted coordinate transformations with
unit Jacobian \eqref{eq:Jacobian-unity}.
Let us, therefore, briefly discuss the issue
of diffeomorphisms.

The authors of Ref.~\cite{Aoki-etal-review-1999}
have given a plausibility argument that
the permutation symmetry over the discrete spacetime points
implies the diffeomorphism invariance of the continuum theory.
But the delicate issue of how simultaneously
the locality appears in the continuum theory
is far from resolved.
We suspect that the nonstandard term \eqref{eq:Seff-nonstandard},
which is local but not fully diffeomorphism invariant, appears due to
some type of interference between the emergence of
locality and the emergence of diffeomorphism invariance
(we are reminded of the appearance of anomalies in
chiral gauge theories).
Anyway, let us have a closer look at how precisely
the surprising term \eqref{eq:Seff-nonstandard} arises
in our calculation.

The actual way how the third term of \eqref{eq:Seff-phi-discrete}
appears is from both the ``internal space''
of each spacetime point individually and the larger
group space of the matrices acting between different
spacetime points.
Specifically, referring to \eqref{eq:AA4123tmp}
in  App.~\ref{subapp:Perturbations} and fixing $k=1$ for convenience,
the origin of the $\phi_{1}^{2}\,\eta_{1}^{2}$
term lies in the $\zeta_{1}$ entries inside the
first $4\times 4$ block on the diagonal
and the origin of the
$\left( \phi_{1}^{2}\,\eta_{2}^{2} + \phi_{2}^{2}\,\eta_{1}^{2} \right)$
and $\left( \phi_{1}^{2}\,\eta_{3}^{2} + \phi_{3}^{2}\,\eta_{1}^{2} \right)$
terms in the $\xi_{12}$ and $\xi_{13}$ entries
``coupling'' the different $4\times 4$ spacetime blocks.

The crucial observation now is that
the way how, for $k \ne l$, the action terms
$\left( \phi_{k}^{2}\,\eta_{l}^{2} + \phi_{l}^{2}\,\eta_{k}^{2} \right)$
arise is essentially the same as
for the action terms $\left(\phi_{k}-\phi_{l}\right)^{2}$;
see the last two paragraphs in App.~\ref{subapp:Perturbations}.  
Both of these terms involve a double sum, each of which
gives a density function $\rho$ for the continuum expression,
as shown by \eqref{eq:double-sum-to-double-integral}.
For the kinetic type terms,
one density function $\rho$ gets absorbed
into the definition of the emerging
inverse metric, as the expression \eqref{eq:emergent-inverse-metric}
makes clear.
But for the mixed $\phi_{k}^{2}\,\eta_{l}^{2}$ terms,
there remains one extra density function $\rho$
in what will become the continuum  Lagrange density
and precisely that $\rho$ gives the
$\sqrt{|g(x)|}$ factor inside the
square brackets on the right-hand side of \eqref{eq:Seff-nonstandard}.

In short, if we can get the double-sum kinetic-type terms
with $\left(\phi_{k}-\phi_{l}\right)^{2}$
in the discrete effective action \eqref{eq:Seff-phi-discrete},
then it is also possible
to get the  double-sum mixed terms with $\phi_{k}^{2}\,\eta_{l}^{2}$.
The first double sum gives the kinetic term
in the continuum action \eqref{eq:Seff-kin-mass},
while the second  double sum gives the nonstandard term
\eqref{eq:Seff-nonstandard}.

In closing, we have a peripheral remark.
We observe, namely, that \eqref{eq:emergent-inverse-metric}
has a direct dependence on the matter function $f$, whereas
\eqref{eq:emergent-root-det} does not.
Starting from the inverse metric
components \eqref{eq:emergent-inverse-metric},
we can, of course, calculate the determinant, but then
all influence of the matter function $f$ must somehow
``average out.'' In any case, taking the
expressions \eqref{eq:emergent-inverse-metric}
and \eqref{eq:emergent-root-det} at face value, it
is clear that the metric determinant appears to play
a special role and it is perhaps not surprising to have
additional $\sqrt{|g(x)|}$ factors turn up in the continuum matter
Lagrange density.\vspace*{-4mm}  

\section{Discussion}
\label{sec:Discussion}

\vspace*{-2mm}  
In a previous paper~\cite{Klinkhamer2022-ext-unimod},
we have explored a cosmological model with a dynamic metric-determinant
field $g(x) \equiv \det g_{\alpha\beta}(x)$,
thereby reducing the allowed coordinate transformations
to those with a unit Jacobian.
Some further new results were presented
in Secs.~\ref{sec:Cosmology-First-model}
and \ref{sec:Cosmology-Second-model} here.

The origin of the nonstandard terms in the matter Lagrange density
with one or more additional factors of $\sqrt{|g(x)|}$,
for example the term from \eqref{eq:Seff-nonstandard},
still needs to be established firmly.
Here, we have presented an explicit calculation based on the
so-called IIB matrix model, which provides a
nonperturbative formulation of superstring theory.
Our basic argument is given 
in Sec.~\ref{subsubsec:Nonstandard-action-term},  
with technical details relegated to App.~\ref{subapp:Perturbations}.  
Considering the effective action of a real
scalar field $\phi(x)$, it appears equally easy to get
the standard kinetic term
$\half\;g^{\alpha\beta}\;
\partial_{\alpha}\phi\,\partial_{\beta}\phi$
in the Lagrange density as nonstandard terms
$\phi^{2}\,\sqrt{|g|}\,\eta^{2}$
or $\eta^{2}\,\sqrt{|g|}\,\eta^{2}$,
for a constant real scalar field $\eta$.
These nonstandard terms are local but only
invariant under restricted coordinate transformations.
The simultaneous appearance of locality and (restricted)
diffeomorphism invariance needs, of course, to be studied further.
The same holds for other approaches to
metric-field extraction from the IIB matrix 
model~\cite{Sakai2019,Steinacker2022,Brahma-etal2022}.

Anyway, the matrix-model calculation of the present paper
alerts us to the possibility that there may be rather
unusual interactions in the effective low-energy theory.
In fact, the nonstandard action term in \eqref{eq:Seff-nonstandard}
corresponds to a type of variable mass square
for the scalar field, where the effective mass square involves
the metric determinant $g(x)$.
As the metric determinant $g(x)$ depends on the environment
through the field equations, there is an obvious
resemblance with the chameleon
scenario~\cite{KhouryWeltman2003-PRL,KhouryWeltman2003-PRD}.
Considering a Lagrange-density
term $\half\,g^2\,m^2\,\phi^2$, for example,
we can perform a simple nonrelativistic analysis using Poisson's
equation and find some possibly interesting behavior,   
but we postpone further discussion to a future publication.

%
%

\begin{appendix}
\section{IIB-matrix-model calculation}
\label{app:IIB-matrix-model-calculation}

\subsection{Emerging spacetime points from master-field matrices}  
\label{subapp:Emerging-spacetime-points}


The bosonic action of the Euclidean IIB matrix
model~\cite{IKKT-1997,Aoki-etal-review-1999} reads
\beq
\label{eq:Sbos}
S_\text{bos}=
-\frac{1}{2}\,\text{Tr}\,
\Big(\big[A^{\alpha},\,A^{\beta} \big]\,\big[ A^{\gamma},\,A^{\delta} \big]\,
\delta_{\gamma\alpha}\,\delta_{\delta\beta}\Big)\,,
\eeq
where the bosonic matrices  $A^{\alpha}$,
with a directional index $\alpha$ running over $\{1,\,\ldots\,,D\}$,
are $N \times N$ traceless Hermitian matrices
and the commutators are defined
by $[B,\,C]\equiv$ $B \cdot C - C \cdot B$
for square ma\-tri\-ces $B$ and $C$ of equal dimension.
The action involves the Kronecker delta $\delta_{\gamma\alpha}$,
which corresponds to a  Euclidean ``metric.''
With matrices $A^{\alpha}$ of the dimension of length
(these matrices will ultimately give the spacetime points $x^{\alpha}$),
the dimension of the action \eqref{eq:Sbos}
is $(\text{length})^{4}$ and the matrix integrals
for the expectation values
have a weight factor $\exp[-S_\text{bos}/\ell^{4}]$
for a model length scale $\ell$.
The genuine IIB matrix model has dimensionality $D=10$.
In this subsection, we keep $D$ general but, elsewhere,
we set $D=4$ when four ``large'' dimensions are considered.

Assume that the
master-field matrices $\widehat{A}^{\,\alpha}$
of the Euclidean IIB matrix model are known and that they are
more or less band-diagonal (with a width $1 < \Delta N \ll N$),
as suggested by exploratory numerical
results in Refs.~\cite{KimNishimuraTsuchiya2012,NishimuraTsuchiya2019,%
Anagnostopoulos-etal-2020} and references therein.
Now, let $K$ be a divisor of $N$, so that
\begin{equation} \label{eq:N-as-product-K-times-n}
N = K\,n\,,
\end{equation}
where both $K$ and $n$ are positive integers.
In the master-field matrices $\widehat{A}^{\,\alpha}$
with  a band-diagonal structure, consider
the $K$ blocks of size $n\times n$ centered on the diagonals
(with $n \gtrsim \Delta N$ and $n \ll N$)
and calculate the averages of the eigenvalues of these blocks.
The obtained averages correspond to the emergent
spacetime points and are denoted
\begin{equation}
\label{eq:xhat-mu-k}
\widehat{x}^{\,\alpha}_{k}\,,
\end{equation}
where $\alpha$ runs over $\{1,\,\ldots ,\,  D\}$ and
$k$  over $\{1,\,\ldots ,\,  K\}$,
with $K$ as given by \eqref{eq:N-as-product-K-times-n}.
Further comments on the extraction procedure
appear in App.~A of Ref.~\cite{Klinkhamer2021-EpiphanyConf}.

The quantities
$\rho(x)$, $r(x,\,y)$, and $f(x)$ entering expressions
\eqref{eq:emergent-inverse-metric-root-det} in the main text
result from the distributions and correlations of the emerging
spacetime points \eqref{eq:xhat-mu-k} and, as regards $f(x)$,
from perturbations of the master-field matrices.

Specifically, the density function $\rho(x)$ and
the density correlation function $r(x,\,y)$ are defined by%
\bsubeqs\label{eq:rho-r-defs}
\beqa \label{eq:rho-def}
\rho(x) &\equiv&
\sum_{k=1}^{K}\;\delta^{(D)} \big(x- \widehat{x}_{k}\big)\,,
\\[2mm]
 \label{eq:r-def}
\langle\,\rho(x)\,\rho(y)\,\rangle
&\equiv&
\langle\,\rho(x)\,\rangle\; \langle\,\rho(y)\,\rangle \; r(x,\,y)\,,
\eeqa
\esubeqs
where $x^{\alpha}$ and $y^{\alpha}$ are
$D$-dimensional continuous (interpolating) coordinates.
The averages $\langle\,\ldots\,\rangle$ in \eqref{eq:r-def}
stand for averaging over different block sizes ($n$)
and over different block positions along the diagonals
of the master-field matrices $\big(\widehat{A}^{\,\alpha}\big)_{kl}$
[note that the block at the beginning of the diagonal has
dimension $n^{\prime} \leq n$ and the block at the end has
dimension $n^{\prime\prime} \leq n$, but there are many
more intermediate $n\times n$ blocks if $K \gg 1$].
In this way, the double sum in  \eqref{eq:Seff-phi-discrete}
is transformed into a double integral
over the continuum spacetime,
\beq
\label{eq:double-sum-to-double-integral}
\sum_{k,\,l}\; s\big(\widehat{x}_{k}-\widehat{x}_{l}\big)\,
\ldots
\;\rightarrow
\int d^{D}x \,\langle\,\rho(x)\,\rangle\;
\int d^{D}y \,\langle\,\rho(y)\,\rangle \;r(x-y)\, s(x-y) \ldots\;,
\eeq
for an arbitrary function $s(x)$.

Finally, the quantity $f(x)$ entering expression
\eqref{eq:emergent-inverse-metric} in the main text
is a localized real function coming from the ``hopping'' term
with $\widetilde{f}\big(\widehat{x}_{k}-\widehat{x}_{l}\big)$
in the discrete scalar effective action \eqref{eq:Seff-phi-discrete}.

\subsection{Perturbations of master-field matrices} 
\label{subapp:Perturbations}

We present here a simple construction to obtain
the third term of the discrete effective action \eqref{eq:Seff-phi-discrete}.
Essentially, this is a variation of the construction method
developed in  App.~A of Ref.~\cite{Klinkhamer2021-master}.
We focus on the  four ``large'' dimensions
(whose appearance may be suggested by exploratory numerical results
in Ref.~\cite{KimNishimuraTsuchiya2012,NishimuraTsuchiya2019,%
Anagnostopoulos-etal-2020}
and references therein) and set $D=4$ in our expressions.

Take, now, the particular matrix sizes
\beq
\label{eq:N-special}
N = 4+ 4\,j +4 \,, \quad j=1,\,2,\,3, \ldots \,.
\eeq
Then, the first and last $4\times 4$ blocks on the diagonal will give
$\phi_{k}^{2}\,\eta_{k}^{2}$ terms in \eqref{eq:Seff-phi-discrete} 
for the smallest and largest values of $k$
and the band diagonal in between (with suitable $4\times 4$
and $2\times 2$ blocks) will give
both $\phi_{k}^{2}\,\eta_{k}^{2}$ terms
and $[\phi_{k}^{2}\,\eta_{k\pm 1}^{2}+\phi_{k\pm 1}^{2}\,\eta_{k}^{2}]$ terms
for intermediate values of $k$. Other far-off entries will give the
$[\phi_{k}^{2}\,\eta_{l}^{2}+\phi_{l}^{2}\,\eta_{k}^{2}]$
terms for $|k-l| \geq 2$. All this will become clearer for the
$j=1$ case to be discussed next.

Indeed, let us focus on the case $N=12$, where the
master-field-type matrices have three
$4\times 4$ blocks on the diagonal, labeled by
$k \in \{1,\,2,\,3\}$.
The basic structure of the perturbed matrices, with
five $2\times 2$ blocks on the diagonal and
two far-off entries (with $\xi_{13}$ and $\xi_{13}^{*}$), 
is then as follows
(with lines added to mark the $4\times 4$ blocks):
\bsubeqs\label{eq:AA4123tmp}
\beqa
\label{eq:AA4tmp}
\hspace*{-10mm}
A^{4}_\text{tmp}&\!\!=\!&
\!\left(
\renewcommand{\arraycolsep}{0.20pc}  
\renewcommand{\arraystretch}{1.25}   
  \begin{array}{cccc|cccc|cccc}
\widehat{x}^{\,4}_{1} & 0 & 0 & 0 & 0 & 0 &
0 & 0 & 0 & 0 & 0 & 0\\
0 & \widehat{x}^{\,4}_{1} & h^{4}\,\zeta_{1} & 0 & 0 & 0 &
0 & 0 & 0 & 0 & 0 & 0\\
0 & h^{4}\,\zeta_{1} & \widehat{x}^{\,4}_{1} + \zeta_{1} & 0 &  0 &
0 & 0 & 0 & 0 & 0 & 0\\
0 & 0 & 0 & \widehat{x}^{\,4}_{1} & \xi_{12} & 0 &
0 & 0 & \xi_{13} & 0 & 0 & 0\\
\hline
0 & 0 & 0 & \xi_{12}^{*} & \widehat{x}^{\,4}_{2} & 0 &
0 & 0 & 0 & 0 & 0 & 0\\
0 & 0 & 0 & 0 & 0 & \widehat{x}^{\,4}_{2} &
i^{4}\,\zeta_{2} & 0 & 0 & 0 & 0 & 0\\
0 & 0 & 0 & 0 & 0 & i^{4}\,\zeta_{2}&
\widehat{x}^{\,4}_{2} + \zeta_{2} & 0 & 0 & 0 & 0 & 0\\
0 & 0 & 0 & 0 & 0 & 0 &
0 & \widehat{x}^{\,4}_{2} & \xi_{23} & 0 & 0 & 0\\
\hline
0 & 0 & 0 & \xi_{13}^{*} & 0 & 0 &
0 & \xi_{23}^{*} & \widehat{x}^{\,4}_{3} & 0 & 0 & 0\\
0 & 0 & 0 & 0 & 0 & 0 &
0 & 0 & 0 & \widehat{x}^{\,4}_{3} & j^{4}\,\zeta_{3} & 0\\
0 & 0 & 0 & 0 & 0 & 0 &
0 & 0 & 0 & j^{4}\,\zeta_{3} & \widehat{x}^{\,4}_{3}+\zeta_{3} & 0\\
0 & 0 & 0 & 0 & 0 & 0 &
0 & 0 & 0 & 0 & 0 & \widehat{x}^{\,4}_{3} \\
 \end{array}
\right)\!
- \Sigma^{4} \id_{12} \,.
\eeqa
where $\id_{12}$ is the $12\times 12$ identity matrix and
$\Sigma^{4}$ makes for tracelessness.
The coefficients $h^{4}$, $i^{4}$, $j^{4}$,
$\zeta_{1}$, $\zeta_{2}$, and $\zeta_{3}$
are real, whereas the  coefficients $\xi_{kl}$ are complex.
The three other matrices are obtained
by straightforward substitutions of the $\widehat{x}^{\,4}_{k}$:
\beqa
\hspace*{-10mm}&&
A^{1}_\text{tmp} =
\nonumber\\
\hspace*{-10mm}&&
A^{4}_\text{tmp} \;\text{with}\;
\{\widehat{x}^{\,4}_{1} \to \widehat{x}^{\,1}_{1},\,
  \widehat{x}^{\,4}_{2} \to \widehat{x}^{\,1}_{2},\,
  \widehat{x}^{\,4}_{3} \to \widehat{x}^{\,1}_{3},\,
  h^{4} \to 0,\,i^{4} \to 0,\,j^{4} \to 0,\,\Sigma^{4}\to\Sigma^{1}\}\,,
\\[2mm]
\hspace*{-10mm}&&
A^{2}_\text{tmp} =
\nonumber\\
\hspace*{-10mm}&&
A^{4}_\text{tmp} \;\text{with}\;
\{\widehat{x}^{\,4}_{1} \to \widehat{x}^{\,2}_{1},\,
  \widehat{x}^{\,4}_{2} \to \widehat{x}^{\,2}_{2},\,
  \widehat{x}^{\,4}_{3} \to \widehat{x}^{\,2}_{3},\,
  h^{4} \to 0,\,i^{4} \to 0,\,j^{4} \to 0,\,\Sigma^{4}\to\Sigma^{2}\}\,,
\\[2mm]
\hspace*{-10mm}&&
A^{3}_\text{tmp} =
\nonumber\\
\hspace*{-10mm}&&
A^{4}_\text{tmp} \;\text{with}\;
\{\widehat{x}^{\,4}_{1} \to \widehat{x}^{\,3}_{1},\,
  \widehat{x}^{\,4}_{2} \to \widehat{x}^{\,3}_{2},\,
  \widehat{x}^{\,4}_{3} \to \widehat{x}^{\,3}_{3},\,
  h^{4} \to 0,\,i^{4} \to 0,\,j^{4} \to 0,\,\Sigma^{4}\to\Sigma^{3}\}\,,
\eeqa
\esubeqs
where, for simplicity,
the $h^{\alpha}$, $i^{\alpha}$, and $j^{\alpha}$ terms
have been set to zero for $\alpha=1,\,2,\,3$.

Next, insert the real perturbations $\phi_{k}$ and $\eta_{k}$
(each with the dimension of length) into the above matrices:
\beqa\label{eq:A4123}
\hspace*{1mm}&&
\{ A^{4},\,A^{1},\,A^{2},\,A^{3}\}=
\nonumber\\
\hspace*{1mm}&&
\{ A^{4}_\text{tmp},\,A^{1}_\text{tmp},\,A^{2}_\text{tmp},\,A^{3}_\text{tmp}\}
\;\text{with}\;
\nonumber\\
\hspace*{1mm}&&
\Big\{
{\zeta_{1}}\rightarrow
    \left( {\eta_{1}}^{2}\,{\phi_{1}}^{2} \right)^{1/4},\;\;
{\zeta_{2}}\rightarrow
    \left( {\eta_{2}}^{2}\,{\phi_{2}}^{2} \right)^{1/4},\;\;
{\zeta_{3}}\rightarrow
\left( {\eta_{3}}^{2}\,{\phi_{3}}^{2} \right)^{1/4},
\nonumber\\
\hspace*{1mm}&&
\xi_{12}\rightarrow
    \widetilde{g}_{12}\,\ell^{-1}\,
    \left( \eta_{2}\,\phi_{1} + i \,\eta_{1}\,\phi_{2} \right) ,\;\;
\xi_{12}^{*}\rightarrow
    \widetilde{g}_{12}\,\ell^{-1}\,
    \left( \eta_{2}\,\phi_{1} - i \,\eta_{1}\,\phi_{2} \right) ,
\nonumber\\
\hspace*{1mm}&&
  \xi_{13}\rightarrow
    \widetilde{g}_{13}\,\ell^{-1}\,
    \left( \eta_{3}\,\phi_{1} + i \,\eta_{1}\,\phi_{3} \right) ,\;\;
  \xi_{13}^{*}\rightarrow
    \widetilde{g}_{13}\,\ell^{-1}\,
    \left( \eta_{3}\,\phi_{1} - i \,\eta_{1}\,\phi_{3} \right) ,
\nonumber\\
\hspace*{1mm}&&
  \xi_{23}\rightarrow
    \widetilde{g}_{23}\,\ell^{-1}\,\left( \eta_{3}\,\phi_{2} +
        i \,\eta_{2}\,\phi_{3} \right) ,\;\;
  \xi_{23}^{*}\rightarrow
    \widetilde{g}_{23}\,\ell^{-1}\,\left( \eta_{3}\,\phi_{2} -
        i \,\eta_{2}\,\phi_{3} \right)
\Big\}     \,,
\eeqa
with real dimensionless coefficients 
$\widetilde{g}_{12},\, \widetilde{g}_{13},\, \widetilde{g}_{23}$
that depend on the differences of the spacetime
points, 
$\widetilde{g}_{12}=\widetilde{g}_{12}(\widehat{x}^{\,\alpha}_{1}-\widehat{x}^{\,\alpha}_{2})$
and similarly for $\widetilde{g}_{13}$ and $\widetilde{g}_{23}$.
Observe that the same coefficients $\zeta_{k}$  
and $\xi_{kl}$ enter all matrices $A^{\alpha}_\text{tmp}$ identically
and precisely these coefficients give the perturbations
by the substitutions \eqref{eq:A4123}; this crucial
point has been emphasized in the second paragraph of
App.~A in Ref.~\cite{Klinkhamer2021-master}.

Evaluating the bosonic action \eqref{eq:Sbos}
for the perturbation matrices \eqref{eq:A4123}
gives
\bsubeqs\label{eq:Sbos-pert-coeff}
\beqa\label{eq:Sbos-pert}
S_\text{bos}^\text{(pert)} &=&
\frac{1}{2} \,\ell^{-2}\, D_{12}\,{\widetilde{g}_{12}}^{\,2}\,
\left( {\eta_{1}}^{2}\,{\phi_{2}}^{2} + {\eta_{2}}^{2}\,{\phi_{1}}^{2}
\right)  +
\frac{1}{2} \,\ell^{-2}\, D_{13}\,{\widetilde{g}_{13}}^{\,2}\,
\left( {\eta_{1}}^{2}\,{\phi_{3}}^{2} + {\eta_{3}}^{2}\,{\phi_{1}}^{2}
       \right) +
\nonumber\\&&
\frac{1}{2}\,\ell^{-2}\, D_{23}\,{\widetilde{g}_{23}}^{\,2}\,
\left( {\eta_{2}}^{2}\,{\phi_{3}}^{2} + {\eta_{3}}^{2}\,{\phi_{2}}^{2}
       \right) +
\nonumber\\&&
\frac{1}{2} \,\left[  3\,\left(h^{4}\right)^{2}\,{\eta_{1}}^{2}\,{\phi_{1}}^{2} +
     3\,\left(i^{4}\right)^{2}\,{\eta_{2}}^{2}\,{\phi_{2}}^{2} +
     3\,\left(j^{4}\right)^{2}\,{\eta_{3}}^{2}\,{\phi_{3}}^{2} \right]\,,
\eeqa
with
\beqa
D_{12}&=&
  3\,\left({\Delta\widehat{x}^{\,1}_{21}}\right)^{2}
+ 3\,\left({\Delta\widehat{x}^{\,2}_{21}}\right)^{2}
+ 3\,\left({\Delta\widehat{x}^{\,3}_{21}}\right)^{2}
- 2\,\Delta\widehat{x}^{\,3}_{21}\,\Delta\widehat{x}^{\,4}_{21}
+ 3\,\left({\Delta\widehat{x}^{\,4}_{21}}\right)^{2} -
\nonumber\\
\hspace*{1mm}&&
  2\,\Delta\widehat{x}^{\,2}_{21}\,\left( \Delta\widehat{x}^{\,3}_{21} +
     \Delta\widehat{x}^{\,4}_{21} \right)  -
  2\,\Delta\widehat{x}^{\,1}_{21}\,\left( \Delta\widehat{x}^{\,2}_{21} +
     \Delta\widehat{x}^{\,3}_{21} + \Delta\widehat{x}^{\,4}_{21} \right)\,,
\eeqa
\beqa
D_{13}&=&
  3\,\left({\Delta\widehat{x}^{\,1}_{21}}\right)^{2}
+ 3\,\left({\Delta\widehat{x}^{\,1}_{32}}\right)^{2}
+ 3\,\left({\Delta\widehat{x}^{\,2}_{21}}\right)^{2}
- 6\,\Delta\widehat{x}^{\,2}_{21}\,\Delta\widehat{x}^{\,2}_{32}
+ 3\,\left({\Delta\widehat{x}^{\,2}_{32}}\right)^{2} -
\nonumber\\&&
  2\,\Delta\widehat{x}^{\,2}_{21}\,\Delta\widehat{x}^{\,3}_{21} +
  2\,\Delta\widehat{x}^{\,2}_{32}\,\Delta\widehat{x}^{\,3}_{21} +
  3\,\left( {\Delta\widehat{x}^{\,3}_{21}}\right)^{2}   +
  2\,\Delta\widehat{x}^{\,2}_{21}\,\Delta\widehat{x}^{\,3}_{32} -
\nonumber\\&&
  2\,\Delta\widehat{x}^{\,2}_{32}\,\Delta\widehat{x}^{\,3}_{32} -
  6\,\Delta\widehat{x}^{\,3}_{21}\,\Delta\widehat{x}^{\,3}_{32} +
  3\,\left( {\Delta\widehat{x}^{\,3}_{32}}\right)^{2}   -
  2\,\Delta\widehat{x}^{\,2}_{21}\,\Delta\widehat{x}^{\,4}_{21} +
\nonumber\\&&
  2\,\Delta\widehat{x}^{\,2}_{32}\,\Delta\widehat{x}^{\,4}_{21} -
  2\,\Delta\widehat{x}^{\,3}_{21}\,\Delta\widehat{x}^{\,4}_{21} +
  2\,\Delta\widehat{x}^{\,3}_{32}\,\Delta\widehat{x}^{\,4}_{21} +
  3\,\left( {\Delta\widehat{x}^{\,4}_{21}}\right)^{2}   +
\nonumber\\&&
  2\,\Delta\widehat{x}^{\,1}_{32}\,\left( \Delta\widehat{x}^{\,2}_{21} -
     \Delta\widehat{x}^{\,2}_{32} + \Delta\widehat{x}^{\,3}_{21} -
     \Delta\widehat{x}^{\,3}_{32} + \Delta\widehat{x}^{\,4}_{21} -
     \Delta\widehat{x}^{\,4}_{32} \right)  -
\nonumber\\&&
  2\,\Delta\widehat{x}^{\,1}_{21}\,\left( 3\,\Delta\widehat{x}^{\,1}_{32} +
     \Delta\widehat{x}^{\,2}_{21} - \Delta\widehat{x}^{\,2}_{32} +
     \Delta\widehat{x}^{\,3}_{21} - \Delta\widehat{x}^{\,3}_{32} +
     \Delta\widehat{x}^{\,4}_{21} - \Delta\widehat{x}^{\,4}_{32} \right)  +
\nonumber\\&&
  2\,\Delta\widehat{x}^{\,2}_{21}\,\Delta\widehat{x}^{\,4}_{32} -
  2\,\Delta\widehat{x}^{\,2}_{32}\,\Delta\widehat{x}^{\,4}_{32} +
  2\,\Delta\widehat{x}^{\,3}_{21}\,\Delta\widehat{x}^{\,4}_{32} -
\nonumber\\&&
  2\,\Delta\widehat{x}^{\,3}_{32}\,\Delta\widehat{x}^{\,4}_{32} -
  6\,\Delta\widehat{x}^{\,4}_{21}\,\Delta\widehat{x}^{\,4}_{32} +
  3\,\left( {\Delta\widehat{x}^{\,4}_{32}}\right)^{2}   \,,
\\[.75mm]
D_{23}&=&
  3\,\left({\Delta\widehat{x}^{\,1}_{32}}\right)^{2}
+ 3\,\left({\Delta\widehat{x}^{\,2}_{32}}\right)^{2}
+ 3\,\left({\Delta\widehat{x}^{\,3}_{32}}\right)^{2}
- 2\,\Delta\widehat{x}^{\,3}_{32}\,\Delta\widehat{x}^{\,4}_{32}
+ 3\,\left({\Delta\widehat{x}^{\,4}_{32}}\right)^{2} -
\nonumber\\&&
  2\,\Delta\widehat{x}^{\,2}_{32}\,\left( \Delta\widehat{x}^{\,3}_{32} +
     \Delta\widehat{x}^{\,4}_{32} \right)  -
  2\,\Delta\widehat{x}^{\,1}_{32}\,\left( \Delta\widehat{x}^{\,2}_{32} +
     \Delta\widehat{x}^{\,3}_{32} + \Delta\widehat{x}^{\,4}_{32} \right)\,,
\\[.75mm]
\Delta\widehat{x}^{\,\alpha}_{21}
&\equiv&
\widehat{x}^{\,\alpha}_{2}-\widehat{x}^{\,\alpha}_{1}\,,
\quad
\Delta\widehat{x}^{\,\alpha}_{32}
\equiv
\widehat{x}^{\,\alpha}_{3}-\widehat{x}^{\,\alpha}_{2}\,.
\eeqa
\esubeqs
The obtained discrete action \eqref{eq:Sbos-pert}  
has the dimension of $(\text{length})^{4}$
and its structure corresponds to the third term on the
right-hand side of \eqref{eq:Seff-phi-discrete}.

We can obtain the first two terms on the
right-hand side of \eqref{eq:Seff-phi-discrete} by enlarging,
for the master-field-type matrices, the $4\times 4$ blocks
on the diagonal to, for example, $6\times 6$ blocks
(corresponding to $N=6+6\,j+6$ with $j\in\mathbb{N}_{+}$).
We have explicitly constructed the
$N=18$ matrices by inserting appropriate $2\times 2$ entries
centered on the diagonal
with the $\kappa_{12}\equiv \widetilde{k}_{12}\,(\phi_{1}-\phi_{2})$
structure as given in App.~A of Ref.~\cite{Klinkhamer2021-master},
by changing the $\zeta_{1}$ replacement
to $\left( {\eta_{1}}^{2}\,{\phi_{1}}^{2} +
\ell^{2}\,{\phi_{1}}^{2} \right)^{1/4}$, and by adding further
appropriate far-off 
terms $\kappa_{13}\equiv \widetilde{k}_{13}\,(\phi_{1}-\phi_{3})$ 
and $\kappa_{13}^{*}=\kappa_{13}$.

For the hopping terms $(\phi_{k}-\phi_{l})^{2}$ with $k\ne l$,
the idea is that, by carefully choosing the rows and columns,
these additional $2\times 2$ entries do
not ``interfere'' with those already present in
\eqref{eq:AA4123tmp}, which were designed
to give the third term on the
right-hand side of \eqref{eq:Seff-phi-discrete}.
The following $4\times 4$ part of the $18\times 18$ matrix $A^{\alpha}$
makes this point clear: 
\beq \label{eq:AA4-18by18-part}
\hspace*{-5mm}A^{\alpha}=
\left(
\renewcommand{\arraycolsep}{0.20pc}  
\renewcommand{\arraystretch}{1.1} 
  \begin{array}{cccccc}
\ddots &  \vdots  &  \vdots  &  \vdots   &   \vdots  &  \udots \\
\ldots &   \widehat{x}^{\,\alpha}_{1} & 0 &
0 & \widetilde{k}_{12}\,(\phi_{1}-\phi_{2})   & \ldots  \\
\ldots &  0  & \widehat{x}^{\,\alpha}_{1} &
\widetilde{g}_{12}\,\left( \phi_{1}\,\eta_{2} + i \,\phi_{2}\,\eta_{1} \right) & 0  &  \ldots \\
\ldots &  0 & \widetilde{g}_{12}\,\left( \phi_{1}\,\eta_{2} - i \,\phi_{2}\,\eta_{1} \right) &
 \widehat{x}^{\,\alpha}_{2} & 0  &  \ldots \\
\ldots &  \widetilde{k}_{12}\,(\phi_{1}-\phi_{2}) & 0 & 0 & \widehat{x}^{\,\alpha}_{2}  &  \ldots \\
\udots &  \vdots  &   \vdots  &  \vdots   &  \vdots   & \ddots  \\
  \end{array}
\right),
\eeq
for which the commutators from \eqref{eq:Sbos} give the action term with
$\left(\phi_{1}^{2}\,\eta_{2}^{2} + \phi_{2}^{2}\,\eta_{1}^{2}\right)$
from the inner $2\times 2$ block
and the action term with $(\phi_{1}-\phi_{2})^{2}$
from the outer $2\times 2$ ``block.''
Both of these $2\times 2$ entries in \eqref{eq:AA4-18by18-part},
the inner one and the outer one, have basically the same structure,
with $\widehat{x}^{\,\alpha}_{1}$ and $\widehat{x}^{\,\alpha}_{2}$
on the diagonal and Hermitian conjugates on the counter-diagonal.

\end{appendix}

\newpage  

\end{document}